\documentclass[aps,prl,preprint,superscriptaddress,longbibliography]{revtex4-1}

\usepackage{graphicx}
\usepackage{xcolor}
\usepackage{amsfonts}
\usepackage{amsmath}
\usepackage{amssymb}
\usepackage{ulem}
\let\oldAA\AA
\renewcommand{\AA}{\text{\normalfont\oldAA}}

\usepackage{setspace}

\def\WS2{WS$_2$}
\def\MoS2{MoS$_2$}

\def\G0W0{G$_0$W$_0$}

\begin{document}
\title{Discovery of interlayer plasmon polaron in graphene/WS$_2$ heterostructures}
\author{S{\o}ren Ulstrup}
\email{ulstrup@phys.au.dk}
\affiliation{Department of Physics and Astronomy, Interdisciplinary Nanoscience Center, Aarhus University, 8000 Aarhus C, Denmark}
\author{Yann in 't Veld}
\affiliation{Institute for Molecules and Materials, Radboud University, 6525 AJ Nijmegen, the Netherlands}
\author{Jill A. Miwa}
\affiliation{Department of Physics and Astronomy, Interdisciplinary Nanoscience Center, Aarhus University, 8000 Aarhus C, Denmark}
\author{Alfred J. H. Jones}
\affiliation{Department of Physics and Astronomy, Interdisciplinary Nanoscience Center, Aarhus University, 8000 Aarhus C, Denmark}
\author{ Kathleen M. McCreary}
\author{Jeremy T. Robinson}
\author{ Berend T. Jonker}
\affiliation{Naval Research laboratory, Washington, D.C. 20375, USA}
\author{Simranjeet Singh}
\affiliation{Department of Physics, Carnegie Mellon University, Pittsburgh, Pennsylvania 15213, USA}
\author{Roland J. Koch}
\affiliation{Advanced Light Source, E. O. Lawrence Berkeley National Laboratory, Berkeley, California 94720, USA}
\author{Eli Rotenberg}
\affiliation{Advanced Light Source, E. O. Lawrence Berkeley National Laboratory, Berkeley, California 94720, USA}
\author{Aaron Bostwick}
\affiliation{Advanced Light Source, E. O. Lawrence Berkeley National Laboratory, Berkeley, California 94720, USA}
\author{Chris Jozwiak}
\affiliation{Advanced Light Source, E. O. Lawrence Berkeley National Laboratory, Berkeley, California 94720, USA}
\author{Malte R{\"o}sner}
\email{m.roesner@science.ru.nl}
\affiliation{Institute for Molecules and Materials, Radboud University, 6525 AJ Nijmegen, the Netherlands}
\author{Jyoti Katoch}
\email{jkatoch@andrew.cmu.edu}
\affiliation{Department of Physics, Carnegie Mellon University, Pittsburgh, Pennsylvania 15213, USA}

\maketitle
\newpage

\textbf{Harnessing electronic excitations involving coherent coupling to bosonic modes is essential for the design and control of emergent phenomena in quantum materials \cite{Basov:2017}.  In situations where charge carriers induce a lattice distortion due to the electron-phonon interaction, the conducting states get ``dressed".  This leads to the formation of polaronic quasiparticles that dramatically impact charge transport, surface reactivity, thermoelectric and optical properties, as observed in a variety of crystals and interfaces composed of polar materials  \cite{Moser:2013,McKeown:2016,Kang:2018,ChenC:2018,Xiang:2023}.  Similarly,  when oscillations of the charge density couple to conduction electrons the more elusive plasmon polaron emerges \cite{Caruso:2015},  which has been detected in electron-doped semiconductors \cite{Riley:2018,Xiaochuan:2021,Caruso:2021}.  However, the exploration of polaronic effects on low energy excitations is still in its infancy in two-dimensional (2D) materials.  Here,  we present the discovery of an interlayer plasmon polaron in heterostructures composed of graphene on top of SL WS$_2$.  By using micro-focused angle-resolved photoemission spectroscopy (microARPES) during \textit{in situ} doping of the top graphene layer, we observe a strong quasiparticle peak accompanied by several carrier density-dependent shake-off replicas around the SL WS$_2$ conduction band minimum (CBM).  Our results are explained by an effective many-body model in terms of a coupling between SL WS$_2$ conduction electrons and graphene plasmon modes.  It is important to take into account the presence of such interlayer collective modes, as they have profound consequences for the electronic and optical properties of heterostructures that are routinely explored in many device architectures involving 2D transition metal dichalcogenides (TMDs) \cite{Britnell:2013,Zihlmann:2018,Arora:2020,Tang:2020,Regan:2020}.}

Sophisticated heterostructure designs involving 2D crystals with pre-defined lattice mismatch and interlayer twist angle have emerged as promising platforms for tailoring potential energy surfaces and excitations in solid state quantum simulators \cite{Tang:2020,Regan:2020}.  While these systems leverage fine-control of complex lattice structures and quantum states, the close proximity of materials may further induce additional interlayer correlation effects \cite{Kennes:2021}.  For example,  in heterostructures composed of graphene and semiconducting TMDs, superlattice bands are generated concomitant with screening-induced band shifts that dictate quasiparticle band alignments and gaps \cite{Ulstrup2:2019,Waldecker:2019,Ulstrup:2020,Xie:2022}.  Intriguingly, recent experiments on twisted bilayer graphene interfaced with SL WSe$_2$ point towards even richer interactions, as the presence of SL WSe$_2$ stabilises superconductivity below the magic twist angle of bilayer graphene \cite{Arora:2020}. In SL WS$_2$ contacted with the topological insulator Bi$_2$Se$_3$, interlayer exciton-phonon bound states have been detected \cite{Hennighausen:2023}. Such observations point to the importance of interlayer collective excitations involving bosonic modes. 
 
\begin{figure*} [t!]
	\begin{center}
		\includegraphics[width=1\textwidth]{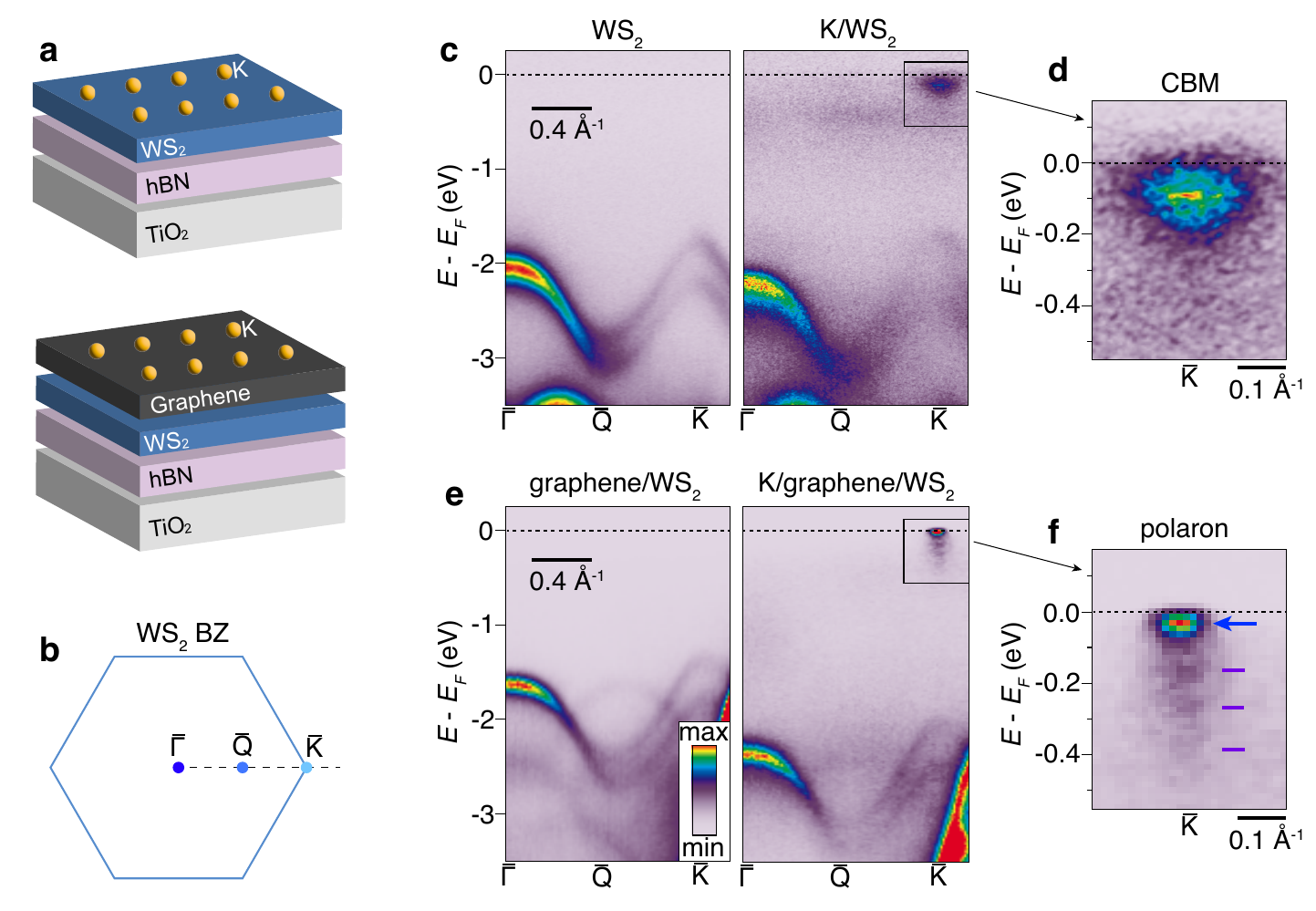}
		\caption{Quasiparticle bands of electron-doped WS$_2$ heterostructures.  \textbf{a,} Layout of systems with doping achieved by deposition of potassium atoms.  \textbf{b,} Brillouin zone (BZ) of SL WS$_2$ with ARPES measurement direction marked by a dashed line. \textbf{c,} ARPES spectra of bare (left panel) and potassium dosed WS$_2$ (right panel) supported on hBN.  The achieved electron density in the strongly doped case is estimated to be $3.0 \cdot 10^{13}$ cm$^{-2}$.  \textbf{d,} Close-up of the CBM region marked in (c).  \textbf{e-f,} Corresponding ARPES spectra for WS$_2$ with graphene on top.  The achieved electron density in the potassium-dosed graphene layer is 4.8 $\cdot$ 10$^{13}$ cm$^{-2}$.  The close-up of the CBM region of WS$_2$ in (f) reveals the formation of a polaron via a sharp quasiparticle peak,  which is demarcated by a blue arrow,  and several shake-off replicas marked by purple ticks.}
		\label{fig:1}
	\end{center}
\end{figure*}

We endeavour to determine how the electronic excitation spectrum of a representative semiconducting SL TMD is affected by a doped graphene overlayer, as is present in a variety of device architectures \cite{Nguyen:2019,Chuang:2014,Cui:2015,Liu:2015,Prisoni:2017,Chee:2019}.  To this end, we focus on SL WS$_2$ as this material exhibits a direct band gap at the $\bar{\mathrm{K}}$-point of the Brillouin zone (BZ) and a large spin-orbit coupling (SOC) induced splitting of the valence bands,  allowing to simultaneously resolve energy- and momentum-dependent electronic excitations around the valence and conduction band extrema using high-resolution ARPES \cite{katoch2018,Hinsche:2017}.  The heterostructures are supported on 10-30 nm thick hBN, which serves two purposes: (i) it replicates the heterostructures that are typically used in transport and optical measurements, and (ii) provides an atomically flat and inert interface that preserves the salient dispersion of SL WS$_2$, since hybridization is strongly suppressed due to the large band gap of hBN \cite{katoch2018}. The entire stack is placed on degenerately-doped TiO$_2$ in order to prevent charging during photoemission. The quasiparticle band structure from the heterostructure is spatially-resolved using microARPES during \textit{in situ} electron doping by depositing potassium atoms on the surface.  In order to determine the effect of the graphene overlayer, we measure two types of heterostructures - one with graphene and one without.  A schematic of our doped heterostructures is presented in Fig.  \ref{fig:1}(a).  Spectra are collected along the $\bar{\mathrm{\Gamma}}$-$\bar{\mathrm{Q}}$-$\bar{\mathrm{K}}$ direction of the SL WS$_2$ BZ,  as sketched in Fig.  \ref{fig:1}(b).

Figure \ref{fig:1}(c) presents ARPES spectra of the effect of strong electron-doping on bare WS$_2$ with potassium atoms deposited directly on the surface.  Before doping,  the expected band structure of SL WS$_2$ is observed with a local valence band maximum (VBM) at $\bar{\mathrm{\Gamma}}$ and a global VBM at $\bar{\mathrm{K}}$, a total gap larger than 2 eV and a SOC splitting of 430 meV in the VBM \cite{Zhu:2011} (see left panel of Fig. \ref{fig:1}(c)).  At an estimated highest electron density of $3.0 \cdot 10^{13}$ cm$^{-2}$, induced by the adsorbed potassium atoms,  the CBM is populated and the shape of the VBM is strongly renormalized,  as observed in the right panel of Fig. \ref{fig:1}(c) and previously reported \cite{katoch2018}. The direct band gap at $\bar{\mathrm{K}}$ is furthermore reduced to $(1.64 \pm 0.02)$ eV (Extended data Fig.  1),  indicating enhanced internal screening.  A detailed view of the CBM region in Fig. \ref{fig:1}(d),  reveals the CBM to be relatively broad with an energy distribution curve (EDC) linewidth of $(0.17 \pm 0.02)$ eV and a momentum distribution curve (MDC) width of $(0.29 \pm 0.02)$ \AA$^{-1}$ (Extended data Fig. 2).

These spectra are contrasted with the situation where a graphene layer is placed on top of WS$_2$ in Fig. \ref{fig:1}(e).  In the undoped case shown in the left panel of Fig. \ref{fig:1}(e),  the bands exhibit the same general features as seen in the left panel of Fig.  \ref{fig:1}(c), although they are noticeably sharper and shifted towards the Fermi energy due to the additional screening of the Coulomb interaction by the graphene \cite{Waldecker:2019}.  Furthermore, a replica of the WS$_2$ local VBM around $\bar{\mathrm{\Gamma}}$ is noticeable close to $\bar{\mathrm{Q}}$ due to the superlattice formed between graphene and WS$_2$ \cite{Ulstrup:2020}. Upon doping graphene to an electron density of about 4.8 $\cdot$ 10$^{13}$ cm$^{-2}$,  the SL WS$_2$ valence band shifts down in energy and the shape of the VBM does not renormalize as in the case of bare WS$_2$ (see right panel of Fig. \ref{fig:1}(e)). The total gap is now $(2.04 \pm 0.02)$ eV (Extended data Fig.  1),  indicating that the non-local Coulomb interaction in WS$_2$ is not fully suppressed.  However,  the CBM region looks dramatically different, as seen by comparing Figs.  \ref{fig:1}(f) and \ref{fig:1}(d).  In the situation with a doped graphene overlayer,  a sharp quasiparticle peak occurs. The peak is accompanied by a series of replica bands towards lower kinetic energy, that are conventionally called shake-off bands. The EDC and MDC linewidths of the main quasiparticle peak are reduced by around a factor of 4, compared to bare K/WS$_2$ (Extended data Fig.  2).  The feature bears resemblance to a Fr{\"o}hlich polaron that is observable in ARPES when the conducting electrons couple strongly to phonons \cite{Moser:2013,McKeown:2016,Xiang:2023,Sio:2023}. 

Density functional theory (DFT) calculations for the K/graphene/WS$_2$ heterostructure (see Methods and Extended data Figs. 3-4) confirm the experimental results which show that the graphene Dirac bands do not strongly hybridize with the WS$_2$ CBM at $\bar{\mathrm{K}}$. As a result, there is only a vanishingly small charge transfer from the strongly K doped graphene layer to the WS$_2$ layer. This explains the experimental observation of strongly doped graphene, accompanied by the small $\bar{\mathrm{K}}$ valley occupation in WS$_2$.  This also explains the absence of VBM renormalization in WS$_2$ covered by graphene, as this only occurs at carrier concentrations larger than 2.0 $\cdot$ 10$^{13}$ cm$^{-2}$ in WS$_2$ \cite{katoch2018}. These DFT calculations, however, do not reproduce the still significant band gap or the shake-off bands,  pointing towards the important role played here by many-body interactions, that are beyond the scope of DFT calculations.

\begin{figure*} [ht!]
	\begin{center}
		\includegraphics[width=0.95\textwidth]{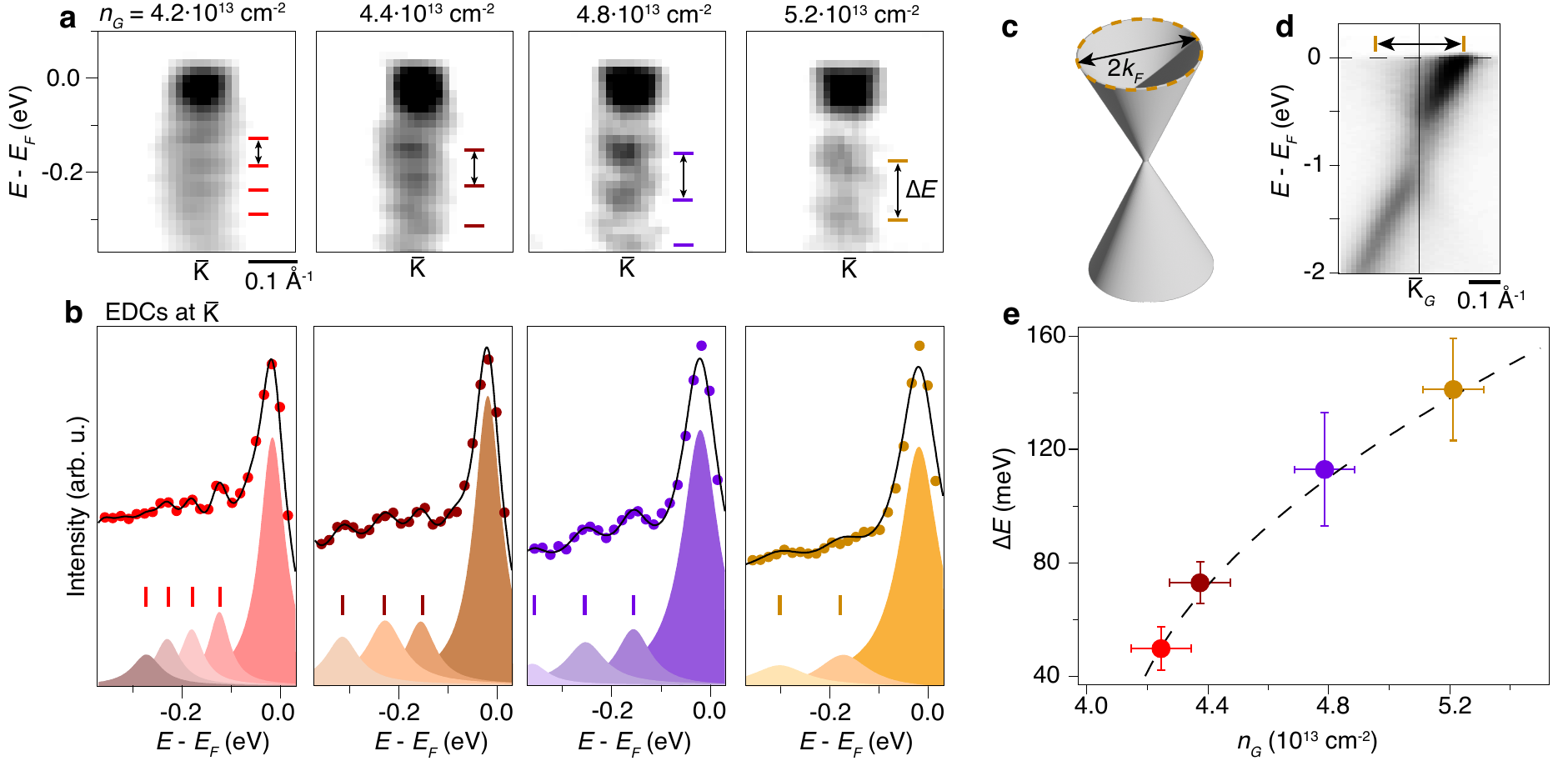}
		\caption{Doping-dependence of shake-off bands.  \textbf{a,} Second-derivative ARPES intensity in the CBM region of potassium dosed  graphene/WS$_2$ at the given electron density in graphene ($n_G$).  Ticks demarcate shake-off bands and the double-headed arrows indicate their energy separation $(\Delta E)$.  \textbf{b,} Energy distribution curves (EDCs) with fits (black curves) to Lorentzian components on a linear background.  Peak components are shown with fitted positions marked by colored ticks.  \textbf{c,} Sketch of graphene Dirac cone and  Fermi surface (dashed circle) measured simultaneously by ARPES at each doping step.   \textbf{d,} ARPES spectrum of potassium dosed graphene on WS$_2$ with diameter of Fermi surface ($2 k_F$) indicated by double-headed arrow. The spectrum is for the maximum achieved doping of graphene of 5.2 $\cdot$ 10$^{13}$ cm$^{-2}$. \textbf{e,} Increase of shake-off energy separation with graphene doping extracted from the analysis.  The dashed line is a fit to a function proportional to $\sqrt{n_G}$.}
		\label{fig:2}
	\end{center}
\end{figure*}

In order to understand the origin of the shake-off bands in the dispersion at $\bar{\mathrm{K}}$ in the graphene/WS$_2$ heterostructure,  we tune the charge carrier density by sequentially increasing the amount of adsorbed potassium on graphene.  After each dosing step we measure both the WS$_2$ conduction band region and the graphene Dirac cone to correlate the evolution of the shake-off bands spectral line shapes with the filling of the Dirac cone.  Second derivative plots of the ARPES intensity are shown in Fig.  \ref{fig:2}(a) to highlight the relatively faint shake-off bands compared to the intense quasiparticle peak for a range of doping where the graphene carrier concentration is varied over a range of $1.0 \cdot 10^{13}$ cm$^{-2}$.  Corresponding EDCs with fits to Lorentzian components are shown in Fig.  \ref{fig:2}(b).  The extraction of the graphene wave vector $k_F$ from measurements of the Dirac cone Fermi surface is illustrated in Fig.  \ref{fig:2}(c) and demonstrated by a measurement of doped graphene on WS$_2$ in Fig.  \ref{fig:2}(d).  The EDC analysis of the shake-off bands as a function of graphene doping reveals the energy separation between shake-off bands increases from $(50 \pm 8)$ meV to $(141 \pm 18)$ meV and that the increase is proportional to $\sqrt{n_G}$,  as shown in Fig.  \ref{fig:2}(e).  Furthermore,  the EDC fits in Fig.  \ref{fig:2}(b) demonstrate that the shake-off band intensity relative to the main quasiparticle peak diminishes with doping.

These observations provide further clues on the origin of the shake-off bands. An internal coupling between WS$_2$ conducting electrons and phonons can be ruled out, because the energy separation of the shake-off bands at high doping exceeds the WS$_2$ phonon bandwidth of 55 meV \cite{Berkdemir:2013}. Given the significant doping of graphene, there are, however, two other bosonic excitations that could be responsible for the shake-off bands in WS$_2$: phonons and plasmons in graphene. In doped graphene there are indeed phonons with energies between 150 and 200 meV with significant electron-phonon coupling. These phonon energies change, however,  only by up to 20 meV upon tuning the electron doping~\cite{novko_dopant-induced_2017,margine_electron-phonon_2016} and can thus be ruled out as the origin for the observed shake-offs. In stark contrast, plasmons in 2D materials are known to be significantly affected by the doping level of the system. Indeed, significant plasmon excitations have been observed in graphene in the regime of doping we are considering \cite{Bostwick2010}. 
Moreover, the doping dependence of the line shape of the shake-off bands is consistent with graphene plasmon excitations coupling to the WS$_2$ conduction electrons \cite{Caruso:2015,Riley:2018,Caruso:2021,Xiaochuan:2021}.  
Taken together, this suggests that the observed feature is an interlayer plasmon polaron with unusually sharp line shapes and well-defined shake-offs, unlike the previously observed plasmonic polarons in electron-doped bulk materials \cite{Riley:2018,Xiaochuan:2021} and in internally doped SL MoS$_2$ \cite{Caruso:2021}.

\begin{figure} [t!]
	\begin{center}
		\includegraphics[width=0.99\textwidth]{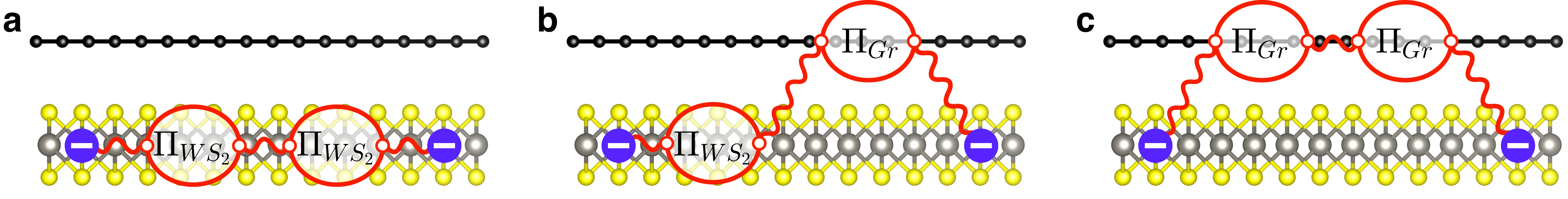}
		\caption{Illustrations of the Coulomb interaction in WS$_2$ and its screening channels in graphene/WS$_2$ heterostructures. Wavy lines and ``bubbles" represent bare Coulomb interactions and polarization processes, respectively. \textbf{a,} Coulomb interaction and screening from WS$_2$ only. \textbf{b,} Illustration of mixed screening channels from WS$_2$ and graphene. \textbf{c,} Coulomb interaction between electrons in WS$_2$ screened by graphene polarization processes only, which couple graphene plasmons to the WS$_2$ Coulomb interaction. Interlayer polarization effects are suppressed due to the vanishingly small hybridization between the WS$_2$ $\bar{\mathrm{K}}$ valley and graphene's Dirac cone.}
		\label{fig:diagram}
	\end{center}
\end{figure}

To theoretically substantiate this interpretation, we use a generic model consisting of two layers, one with a parabolic electronic spectrum mimicking the occupied \WS2 $\bar{\mathrm{K}}$-valley and one with the conventional Dirac spectrum of graphene. As justified by our DFT calculations, we assume that the two layers are electronically decoupled, such that the only coupling between them is the long-range Coulomb interaction. 
Based on this model, we perform random phase approximation (RPA) calculations to evaluate the momentum-dependent and dynamically screened density-density Coulomb interaction matrix $\mathbf{W}_q(\omega)$ in the WS$_2$/graphene layer basis. In Fig.~\ref{fig:diagram} we depict the possible screening channels to the Coulomb interaction between electrons within the WS$_2$ layer, which can be categorized into WS$_2$ screening only [Fig.~\ref{fig:diagram}(a)], mixed screening from WS$_2$ and graphene [Fig.~\ref{fig:diagram}(b)], and pure graphene screening [Fig.~\ref{fig:diagram}(c)].  
We subsequently use $\mathbf{W}_q(\omega)$ within a \G0W0 framework to calculate the interacting spectral function within the effective \WS2 $\bar{\mathrm{K}}$-valley.

\begin{figure*} [t!]
	\begin{center}
		\includegraphics[width=1\textwidth]{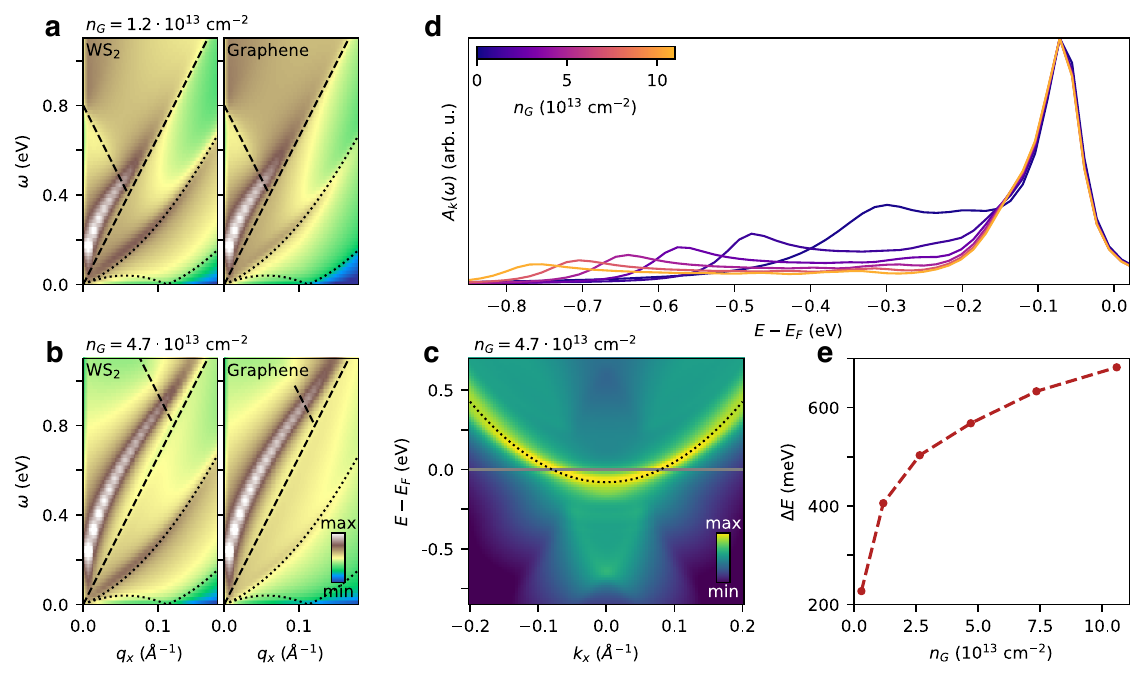}
		\caption{Theoretical \G0W0 results. \textbf{a-b,} Spectral function $B_q(\omega)$ of the Coulomb interaction in the WS$_2$ (left) and graphene (right) channels, for two different graphene occupations $n_G$. Dotted and dashed lines indicate the edges of the electron-hole continua of WS$_2$ and graphene, respectively.  \textbf{c,} Momentum resolved WS$_2$ normal state spectral function $A_k(\omega)$. The dotted black line represents the bare WS$_2$ dispersion.  \textbf{d,} EDCs of the WS$_2$ normal state spectral function at $\bar{\mathrm{K}}$ for a variety of graphene occupations.  \textbf{e,} Energy splitting $\Delta E$ between the WS$_2$ quasiparticle peak and the plasmon polaron peak as a function of graphene occupation.}
		\label{fig:3}
	\end{center}
\end{figure*}

In Figs. \ref{fig:3}(a) and \ref{fig:3}(b) we show elements of $\mathbf{B}_q(\omega) = -\pi^{-1}\text{Im}\left[\mathbf{W}_q(\omega)\right]$, which is proportional to the momentum resolved electron energy loss spectrum EELS$_q(\omega)$, within the \WS2 (left panels) and graphene (right panels) layers for two different graphene carrier concentrations. Within the graphene layer, we find a strong plasmonic resonance with the characteristic $\sqrt{q}$-like dispersion \cite{koppens_graphene_2011,grigorenko_graphene_2012}, which is damped as soon as it enters graphene's particle-hole continuum indicated by black dashed lines.
Upon doping graphene, the plasmon dispersion shifts to higher energies, but maintains its $\sqrt{q}$-like dispersion. 
Within the \WS2 layer, we find two distinct branches, corresponding to two different plasmonic modes. 
The lower branch closely follows the original \WS2 plasmon dispersion which we would find upon excluding interlayer Coulomb coupling from processes as those illustrated in Figs.~\ref{fig:diagram}(b)-(c), such that we identify this branch as a \WS2 intralayer plasmon dominated one.
The upper branch arises from the interlayer Coulomb coupling and closely follows the graphene plasmon dispersion. The graphene plasmon dispersion thus strongly couples into the screened Coulomb interaction within the \WS2 layer. This is a consequence of the vastly different electron densities in the two layers and the enhanced graphene Landau damping region, which overlaps the undamped $(q,\omega)$ area of the \WS2 polarization. This finally leads to a strong suppression of the \WS2 plasmonic features within both layers and is a result of the significantly larger Fermi velocity in graphene.
As a consequence, the dynamics of the \WS2 Coulomb interaction is dominated by screening from the graphene layer as illustrated in Fig.~\ref{fig:diagram}(c), and is therefore highly dependent on the graphene doping through its plasmon.

In Fig. \ref{fig:3}(c) we show the resulting dressed spectral function $A_k(\omega) = -\pi^{-1}\text{Im}\left[G_k(\omega)\right]$ within the effective \WS2 $\bar{\mathrm{K}}$-valley for a graphene doping of $n_G = 4.7 \cdot 10^{13}$\,cm$^{-2}$. In qualitative agreement with the ARPES spectrum, we find a pronounced plasmon polaron band below the original CBM separated by an energy $\Delta E$. In Fig. \ref{fig:3}(d) we further show EDCs of the spectral function $A_k(\omega)$ at the center of the conduction band for a variety of graphene occupations. In all of these linecuts we can identify the main quasiparticle around $E - E_F \approx -0.08$\,eV, and the plasmon polaron bands at lower energies. As the graphene occupation increases, the plasmon polaron band shifts to lower energies, increases its separation from the main quasiparticle peak, and loses intensity. The energy splitting $\Delta E$ is plotted as a function of graphene doping $n_G$ in Fig. \ref{fig:3}(e), where we find the same $\sqrt{n_G}$-like behaviour as observed in our ARPES measurement.  This trend is fully driven by the graphene plasmon and is unaffected by the WS$_2$ polarization as discussed in the Extended Material.

The higher order shake-off bands are missing as a result of the \G0W0 approximation and can only be described by taking vertex corrections into account~\cite{aryasetiawan_multiple_1996,guzzo_valence_2011,Caruso:2021}.  Missing vertex corrections~\cite{aryasetiawan_multiple_1996, guzzo_valence_2011,caruso_gw_2016} and underestimated external (substrate) screening (see Extended Material) are reasons for the theoretically estimated $\Delta E$ being larger than the experimental ones. To obtain a  quantitative agreement, we would additionally need to derive all model parameters from first principle calculations based on the full non-commensurate supercell heterostructures and taking SOC effects into account.  However, the details of the parameters do not influence the qualitative behaviour of the theoretical results as shown in Extended data Fig.  5. The qualitative agreement between our theoretical and experimental results is thus striking, such that we conclude that the observed shake-off bands in \WS2 are induced by the graphene plasmon, forming interlayer plasmonic polaron satellites in \WS2. 

The impact of this finding could be far-reaching, as interfaces between graphene and TMDs have been exploited in various ways: to induce large spin-orbital proximity effects \cite{Zihlmann:2018}, for the stabilization of superconductivity below magic angle twists in bilayer graphene interfaced with WSe$_2$ \cite{Arora:2020}, or for charge carrier control of Wigner crystallization and realizations of discrete Mott states in dual-gated TMD heterobilayers contacted with graphite \cite{Tang:2020,Regan:2020}. Our observation of interlayer polaronic quasiparticles induced by adding charge to a contacting graphene layer will thus be important to consider in the interpretation and modelling of device measurements.  Further experiments will be required to evaluate their impact on the optoelectronic properties and band engineering of heterostructures as well as their utility for ultrathin photonics and plasmonic devices.

\section{acknowledgement}
Y.I.V. and M.R. thank G. Ganzevoort for useful discussions. J.K. acknowledges funding from the U.S. Department Office of Science, Office of Basic Sciences, of the U.S. Department of Energy under Award No. DE-SC0020323 as well as partial support by the Center for Emergent Materials, an NSF MRSEC, under award number DMR-2011876. S.U.  acknowledges funding from the Danish Council for Independent Research, Natural Sciences under the Sapere Aude program (Grant No. DFF-9064-00057B) and from the Novo Nordisk Foundation (Grant NNF22OC0079960). J.A.M acknowledges funding from the Danish Council for Independent Research, Natural Sciences under the Sapere Aude program (Grant No. DFF-6108-00409) and the Aarhus University Research Foundation. S.S. acknowledges the support from National Science Foundation under grant DMR-2210510 and the Center for Emergent Materials, an NSF MRSEC, under award number DMR-2011876. K.M.M, J.T.R. and B.T.J. acknowledge support from core programs at the Naval Research Laboratory. Y.I.V and M.R. acknowledge support from the Dutch Research Council (NWO) via the “TOPCORE” consortium. M.R. acknowledges partial support by the European Commission’s Horizon 2020 RISE program Hydrotronics (Grant No. 873028).

\section{Author information}

Correspondence and requests for materials should be addressed to:

S.U. (ulstrup@phys.au.dk).

M.R. (m.roesner@science.ru.nl).

J. K. (jkatoch@andrew.cmu.edu).

\section{Supporting Information}

\subsection{Fabrication of heterostructures}

First, bulk hBN crystals (commercial crystal from HQ Graphene) were exfoliated onto 0.5 wt\% Nb-doped rutile TiO$_2$(100) substrate (Shinkosha Co., Ltd) using scotch tape to obtain 10-30 nm thick hBN flakes. Next, we transferred chemical vapor deposition (CVD) grown SL WS$_2$ onto a selected thin hBN flake using a thin polycarbonate film on a polydimethylsiloxane stamp using a custom-built transfer tool. This was followed by transfer of CVD graphene on top of the WS$_2$/hBN stack. The details of the CVD growth of SL WS$_2$ and graphene as well as transfer process were previously provided in Ref. \citenum{Ulstrup:2020}. After each transfer process, the sample surface was cleaned by annealing in ultrahigh vacuum (UHV) at 150 $^{\circ}$C for 15 mins to remove any unwanted residues or adsorbates from the surface.

\subsection{Micro-focused angle-resolved photoemission spectroscopy}

The photoemission experiments were carried out in the microARPES end-station of the MAESTRO facility at the Advanced Light Source.  Samples were transported through air and given a 1 hour anneal at 500 K in the end-station prior to measurements.  The base pressure of the system was better than $5 \cdot 10^{-11}$ mbar and the samples were kept at a temperature of 78 K throughout the measurements. 

Energy- and momentum-resolved photoemission spectra were measured using a Scienta R4000 hemispherical electron analyser with custom-made deflectors.  All samples were aligned with the $\bar{\mathrm{\Gamma}} - \bar{\mathrm{K}}$ direction of the WS$_2$ Brillouin zone (BZ) aligned along the slit of the analyser.  Measurements on WS$_2$ samples without a graphene overlayer were performed with a photon energy of 145 eV, while measurements on samples with a graphene overlayer were done with a photon energy of 80 eV.  These energies were chosen on the basis of photon energy scans revealing the optimum matrix elements for clearly resolving the WS$_2$ and graphene band structures.  The photon beam was focused to a spot-size with a lateral diameter of approximately 10 $\mu$m using Kirkpatrick-Baez (KB) mirrors.

Electron-doping of samples was achieved by depositing potassium (K) from a SAES getter source \textit{in situ}.  Each dose had a duration of 40 s. After each dose,  the $\bar{\mathrm{\Gamma}} - \bar{\mathrm{K}}$ cut of WS$_2$ was acquired for 5 minutes followed by a measurement around the Dirac point of graphene for 3 minutes.  Efficient switching between these two cut directions was achieved using the deflector capability of the analyser,  such that all measurements could be done with the sample position held fixed.  In WS$_2$ without a graphene overlayer,  the carrier concentration in WS$_2$ was estimated using the Luttinger theorem via the Fermi surface area enclosed by the WS$_2$ conduction band.  In the samples with a graphene overlayer,  we determined the doping of graphene by directly measuring $k_F$, as shown in Figs.  2(c) and 2(d) in the main paper,  and use the relation $n_G = k_F^2 /\pi$.

The second derivative plots of the ARPES intensity shown in Fig. 2(a) of the main paper were obtained using the method described in Ref.  \citenum{CurvM:2011} and merely used as a tool to visualize the data.  Analysis of energy and momentum distribution curves were always performed on the raw ARPES intensity.

A total of 3 samples were studied, which were a bare WS$_2$ and two graphene/WS$_2$ heterostructures on separate TiO$_2$ wafers such that fresh doping experiments could be performed on all samples.  The two graphene/WS$_2$ heterostructures exhibited twist angles of (7.5 $\pm$ 0.3)$^{\circ}$ and (18.1 $\pm$ 0.3)$^{\circ}$ between graphene and WS$_2$, as determined from the BZ orientations in the ARPES measurements.  We found identical behaviors with doping and formation of polarons in the two heterostructures,  confirming the reproducibility of our results.

\subsection{Density Functional Theory Calculations} 

To study the hybridization and the possible charge transfer between the graphene and \WS2 layers, we performed density functional theory (DFT) calculations using a $4 \times 4$ \WS2 / $5 \times 5$ graphene supercell with K doping, as indicated in Fig.~\ref{fig:structure}. The supercell height has been fixed to about $26$ \AA\ to suppress unwanted wavefunction overlap between adjacent supercells. The \WS2 lattice constant has been fixed to its experimental value of $3.184$ \AA\, while the graphene lattice constant has been strained by about $3\%$ to $2.547$ \AA\ to obtain a commensurable heterostructure. The graphene-\WS2 interlayer separation has been set to previously reported $3.44$ \AA~\cite{hernangomez-perez_charge_2023} and the K-graphene distance has been optimized in DFT yielding $2.642$ \AA\ in the out-of-plane direction. All calculations were performed within the Vienna Ab initio Simulation Package (VASP)~\cite{kresse_ab_1993,kresse_efficient_1996} utilizing the projector-augmented wave (PAW)~\cite{kresse_ultrasoft_1999,blochl_projector_1994} formalism within the PBE~\cite{perdew_generalized_1996} generalized-gradient approximation (GGA) using $12 \times 12 \times 1$ $k$ point grids and an energy cut-off of $400\,$eV.

In Fig.~\ref{fig:dft} we show the resulting unfolded band structure (without SOC effects) together with the pristine \WS2 band structure (including SOC effects) following the approach from Ref. \cite{popescu_extracting_2012} as implemented in \cite{qijingzhengvaspbandunfolding_2023}. From this we can clearly see that in the heterostructure new states in the gap of \WS2 arise, which we identify as graphene bands.  Due to unfolding (matrix element) effects, the second linear band forming graphene's Dirac cone is not visible.  Upon unfolding to the primitive graphene structure, the Dirac point becomes visible (right panel) showing a graphene Fermi energy of about $0.6$ eV in good agreement with the experimentally achieved range. In the upmost valence states around the $\bar{\mathrm{K}}$-point, we see that the graphene and \WS2 bands hybridize similar to reported band structures on undoped graphene/\WS2 \cite{hofmann_link_2023,hernangomez-perez_charge_2023}. In the conduction band region, we however see that graphene states are far from the $\bar{\mathrm{K}}$-valley, such that hybridization between graphene $p_z$ and W $d_{z^2}$ orbitals (which are dominating the $\bar{\mathrm{K}}$-valley) is almost completely suppressed. As a result, there is negligible charge transfer from graphene to \WS2, so that primarily graphene is doped by potassium. This is fully in line with our experimental results.

\subsection{\G0W0 calculations}

We can consider the \WS2 layer to be weakly doped and the $\bar{\mathrm{K}}$-valley to be electronically decoupled from the strongly doped graphene layer. Consequently, we can approximate the \WS2 $\bar{\mathrm{K}}$-valley dispersion by a single-band effective mass model $\xi_k = k^2 / (2 m^*) - \mu_{{\text{WS}}_2}$, with $\mu_{{\text{WS}}_2}$ the \WS2 chemical potential and neglecting SOC effects. Due to the electronic decoupling, the two layers only interact via the long-range Coulomb interaction.

The WS$_2$ effective-mass was set to $m^* = 0.3 m_e$. The interlayer distance, WS$_2$ and graphene unit-cell areas and internal dielectric constants were fixed to $d = 3$\,\AA, $A_{{\text{WS}}_2} = 8.79$\,\AA$^2$, $A_{\text{Gr}} = 5.27$\,\AA$^2$, $\varepsilon_{{\text{WS}}_2} = 8.57$ and $\varepsilon_{\text{Gr}} = 2.0$, respectively. Additionally, the external dielectric constant was set to $\varepsilon_{ext} = 3$, close to the value of hBN, and the material heights were set to $h_{{\text{WS}}_2} = 6.16$\,\AA\, and $h_{\text{Gr}} = 3.35$\,\AA\, for \WS2 and graphene, respectively, in accordance with Refs.~\cite{rosner_wannier_2015,steinke_coulomb-engineered_2020}. We set the WS$_2$ chemical potential to $\mu_{{\text{WS}}_2} = 0.04$ eV such that WS$_2$ had an occupation of $n = 1.7 \cdot 10^{13}$\,cm$^{-2}$ and varied the graphene chemical potential to simulate various graphene occupations. The \G0W0 calculations were performed with the TRIQS~\cite{parcollet_triqs_2015} and TPRF~\cite{wentzell_triqstprf_2022} packages, using $300 \times 300$ momentum meshes and $500$ linearly spaced frequency points from $-4.0$\,eV to $4.1$\,eV, with a broadening of $\delta = 0.02$ eV.\\

In order to describe the non-local and dynamic screening of the heterostructure, we calculated the screened Coulomb matrix $\mathbf{W}_q(\omega)$ in the random phase approximation (RPA), neglecting non-density-density terms. Due to the negligible interlayer hybridization, the density-density RPA polarization $\mathbf{\Pi}_q(\omega)$ is approximately block diagonal,
\begin{equation}
\mathbf{\Pi}_q(\omega) =
\begin{bmatrix}
  \Pi^{\text{WS2}}_q(\omega) & 0\\
  0 & \Pi^{\text{Gr}}_q(\omega)
\end{bmatrix},
\end{equation}
such that for both the \WS2 block $\Pi^{\text{WS2}}_q(\omega)$ and the graphene block $\Pi^{\text{Gr}}_q(\omega)$ we can independently use analytic zero-temperature expressions for a quadratic~\cite{giuliani_quantum_2005} and a Dirac dispersion~\cite{wunsch_dynamical_2006,hwang_dielectric_2007}, respectively.

The bare density-density Coulomb matrix elements were approximated by
\begin{equation}
\mathbf{V}_q = V_q
\begin{bmatrix}
  1  & e^{-q d}\\
  e^{-q d} & 1
\end{bmatrix},
\end{equation}
where $d$ is the interlayer distance and $V_q$ is given by $V_q = 2 \pi e^2 / (A \varepsilon_q q)$, with $A = (A_{{\text{WS}}_2} + A_{\text{Gr}})/2$ the layer-averaged unit cell area and $e$ the electron charge. $\varepsilon_q$ is a dielectric function capturing the neglected non-metallic screening channels and is given by~\cite{keldysh_coulomb_1979}
\begin{equation}
    \varepsilon(q) = \varepsilon_{int} \frac
    {1 - \tilde{\varepsilon}^2 e^{-2 q h}}
    {1 + 2 \tilde{\varepsilon} e^{-qh} + \tilde{\varepsilon}^2 e^{-2 q h}},
\end{equation}
with $\tilde{\varepsilon} = \left(\varepsilon_{int} - \varepsilon_{ext}\right) / \left(\varepsilon_{int} + \varepsilon_{ext}\right)$ and $h = h_{{\text{WS}}_2} + h_{\text{Gr}}$ the total thickness of the heterostructure. The internal dielectric constant is set to the average value of \WS2 and graphene, i.e., $\varepsilon_{int} = (\varepsilon_{{\text{WS}}_2} + \varepsilon_{\text{Gr}})/2$.
The screened density-density Coulomb interaction is finally given as the matrix product $\mathbf{W}_q(\omega) = \left(\mathbf{I} - \mathbf{V}_q \mathbf{\Pi}_q(\omega)\right)^{-1} \mathbf{V}_q$, with $\mathbf{I}$ the identity matrix.

To explain the experimental observation of plasmon polaron bands in the doped graphene/\WS2 heterostructure we need to theoretically describe the effect of the non-local and dynamic Coulomb interaction on the \WS2 normal state, for which we performed G$_0$W$_0$ calculations. The negligible hybridization between the layers renders both the bare electron propagator $\mathbf{G}^{(0)}$ and the G$_0$W$_0$ self-energy $\mathbf{\Sigma}$ approximately block diagonal. As a consequence, one only requires information about the screened Coulomb interaction within the \WS2 layer, $W^{{\text{WS}}_2}_q(\omega)$, to determine the renormalization of the \WS2 normal state. The self-energy is therefore given by 
\begin{equation}
    \Sigma^{{{\text{WS}}_2}}_{k}(i\omega_n) = -\frac{1}{\beta} \sum_{q, m} G^{(0), {{\text{WS}}_2}}_{k+q}(i\omega_n + i\omega_m) W^{{{\text{WS}}_2}}_q(i\omega_m),
\end{equation}
where $G^{(0), {{\text{WS}}_2}}_{k}(i\omega_n) = 1 / (i\omega_n - \xi_k)$. 
In spectral representation the sum over discrete Matsubara frequencies can be performed analytically, such that on the real-frequency axis
\begin{equation}
    \Sigma^{{{\text{WS}}_2}}_{k}(\omega) = 
        -\sum_{q} V_q n_F(\xi_{k+q})
        + \sum_{q} \int d\omega' B_q(\omega')
        \frac{n_B(\omega') + n_F(\xi_{k+q})}{\omega + i\delta + \omega' - \xi_{k+q}},
\end{equation}
where $B_q(\omega) = -\pi^{-1}\text{Im}\left(W^{{{\text{WS}}_2}}_q(\omega) - V_q\right)$ and $n_F$ and $n_B$ are the Fermi-Dirac and Bose-Einstein distribution functions, respectively. To compare to ARPES measurements we calculated the \WS2 spectral-function as $A^{{{\text{WS}}_2}}_k(\omega) = -\pi^{-1}\text{Im}\left( G^{{{\text{WS}}_2}}_k(\omega) \right)$, where $G^{{{\text{WS}}_2}}_k(\omega) = 1/\left(\omega + i\delta - \xi_k - \Sigma^{{{\text{WS}}_2}}_k(\omega)\right)$.\\

In Fig.~\ref{fig:ex5} we show the spectral function of WS$_2$ and the resulting $\Delta E$ (as insets) for various further model parameters. From these additional calculations, we understand that the WS$_2$ polarization does not play a significant role for the interlayer plasmon polaron formation and that the effective distance, effective mass, and effective external screening only quantitatively affect the shake-off band splitting.  Note that increasing the effective external screening (as resulting from the substrate) from $\varepsilon_{ext}=3$ to $\varepsilon_{ext}=6$ reduces $\Delta E$ significantly.

\begin{figure*} [t!]
	\begin{center}
		\includegraphics[width=0.8\textwidth]{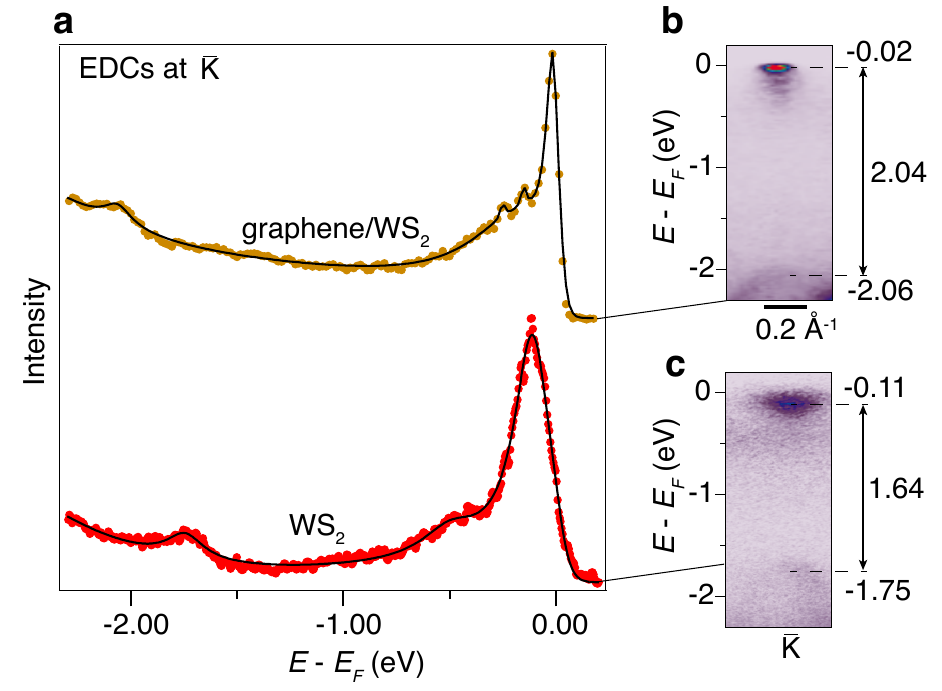}
		\caption{Extraction of band gaps from ARPES.  \textbf{a,} Energy distribution curves (EDCs) at $\bar{\mathrm{K}}$ in the heavily doped situations shown in Figs.  1(c) and 1(e) of the main paper.  The black curves represent fits to a modelled spectral function with several Lorentzian peaks on a linear background and including a Fermi-Dirac cut-off.  \textbf{b-c,} ARPES spectra around $\bar{\mathrm{K}}$,  displaying the VBM and CBM of SL WS$_2$ (b) with and (c) without a graphene overlayer. VBM and CBM peak positions obtained from the EDC analysis in (a) and the resulting band gaps are stated in units of eV.  The error bars are $\pm 0.02$ eV.}
		\label{fig:ex1}
	\end{center}
\end{figure*}

\begin{figure*} [t!]
	\begin{center}
		\includegraphics[width=1\textwidth]{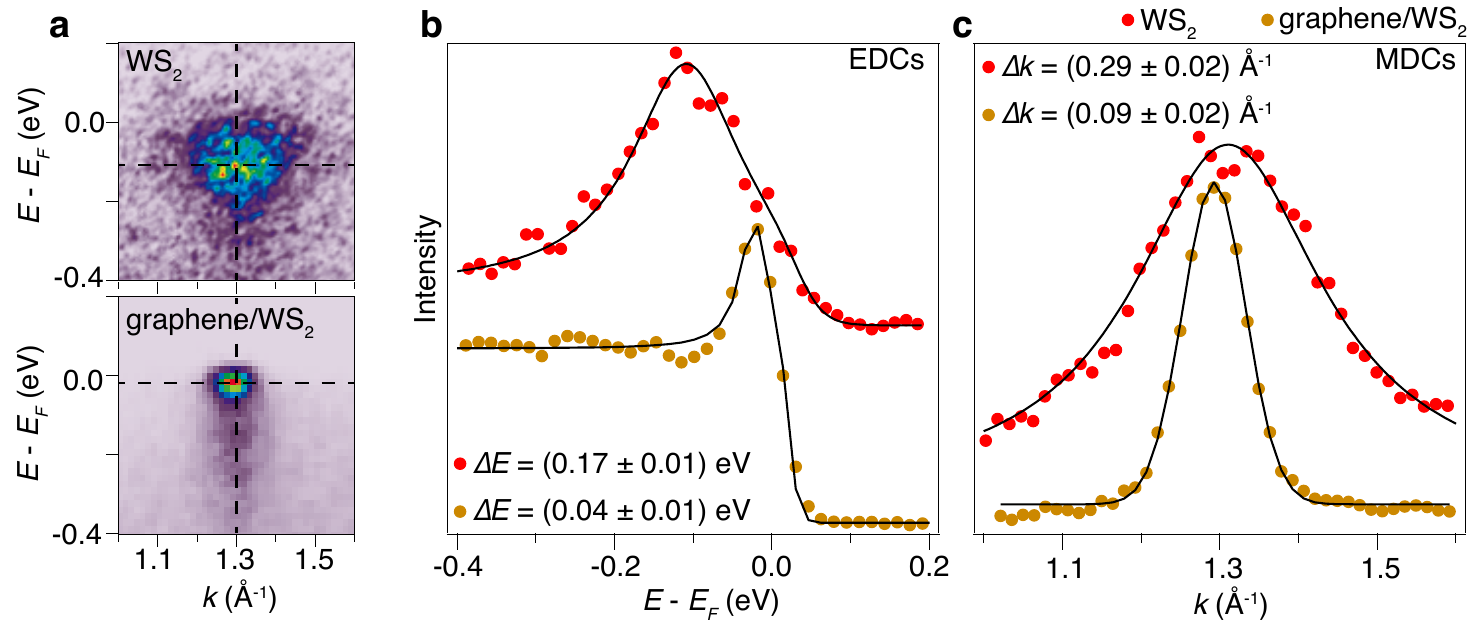}
		\caption{ARPES linewidths of conduction band states.  \textbf{a,} ARPES spectra in the conduction band region of WS$_2$ and graphene/WS$_2$,  corresponding to the data shown in Figs. 1(d) and 1(f).  \textbf{b,} EDCs (markers) extracted along the vertical dashed lines in (a) with fits (black curves) to a single Lorentzian peak multiplied by a Fermi-Dirac function.  The linewidth of the Lorentzian peak is stated as $\Delta E$.  \textbf{c,} Momentum distribution curves (MDCs) extracted at the EDC peak energies marked by horizontal dashed lines in (a) and fits to a Lorentzian peak convoluted with a Gaussian. The resulting momentum linewidth is stated as $\Delta k$.}
		\label{fig:ex2}
	\end{center}
\end{figure*}

\begin{figure*} [t!]
	\begin{center}
		\includegraphics[width=1\textwidth]{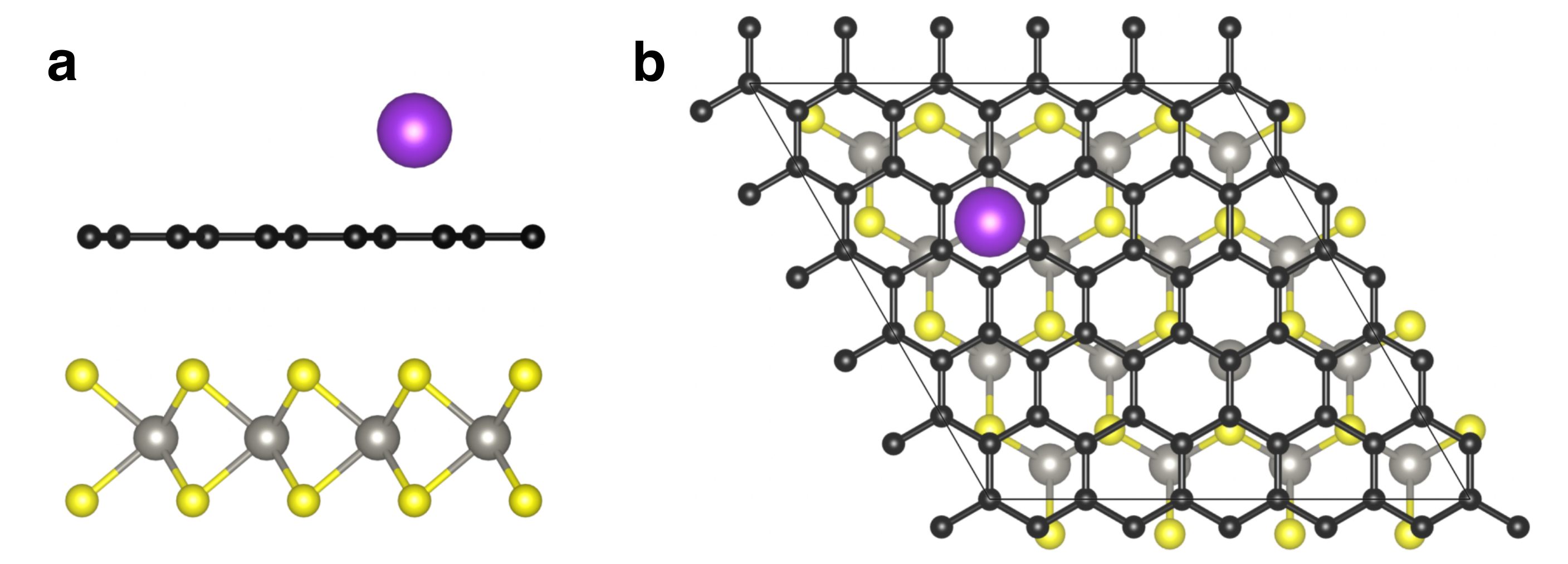}
		\caption{DFT heterostructure model. \textbf{a-b,} View from the (a) side and (b) top of the utilized $4 \times 4$ \WS2 / $5 \times 5$ graphene supercell with K doping. }
		\label{fig:structure}
	\end{center}
\end{figure*}

\begin{figure*} [t!]
	\begin{center}
		\includegraphics[width=1\textwidth]{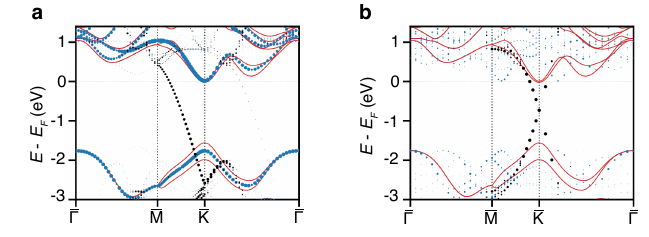}
		\caption{DFT electronic structure.  \textbf{a-b,} Band structure of \WS2 with SOC (red line) plotted together with the unfolded band structure without SOC (dots) in the (a) WS$_2$ and (b) graphene primitive BZs.  Blue and black dots represent W and C weights.}
		\label{fig:dft}
	\end{center}
\end{figure*}

\begin{figure*} [t!]
	\begin{center}
		\includegraphics[width=1\textwidth]{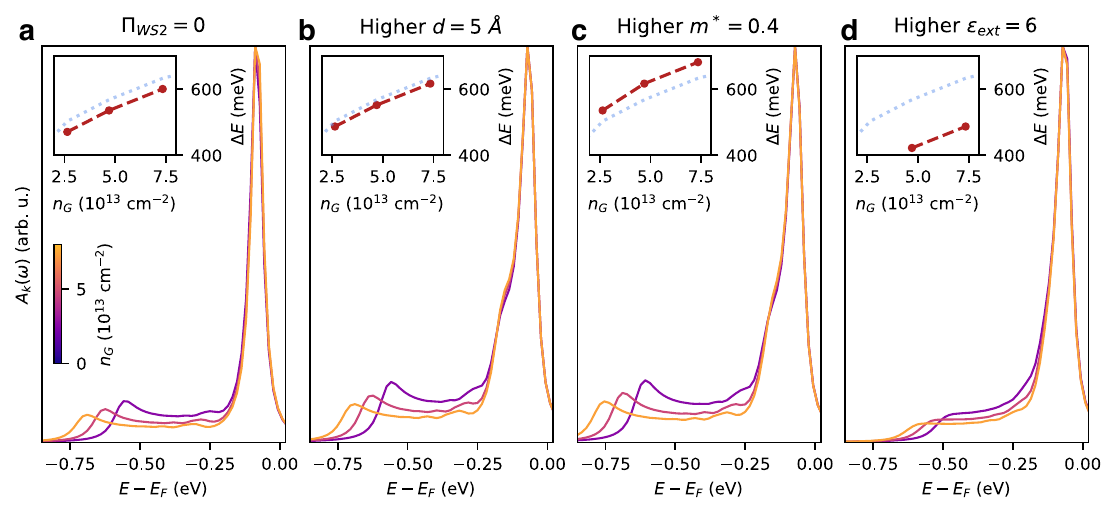}
		\caption{\WS2 normal state spectral function $A_k(\omega)$ at $\bar{\mathrm{K}}$ obtained from \G0W0 calculations for a variety of graphene occupations. All model parameters are the same as those used in the main text, except for one change per figure. \textbf{a,} the \WS2 polarization is set to 0. \textbf{b,} the interlayer distance is increased to $d = 5$\,\AA. \textbf{c,} the effective mass is increased to $m^* = 0.4$. \textbf{d,} the external dielectric constant is increased to $\varepsilon_{ext} = 6$. The insets show the energy splitting $\Delta E$ between the WS$_2$ quasi-particle peak and the plasmon polaron peak as a function of graphene occupation, for the changed model parameters (red dashed) and for the model parameters used in the main text (blue dotted). }
		\label{fig:ex5}
	\end{center}
\end{figure*}


\newpage
\clearpage

\begin{thebibliography}{58}%
\makeatletter
\providecommand \@ifxundefined [1]{%
 \@ifx{#1\undefined}
}%
\providecommand \@ifnum [1]{%
 \ifnum #1\expandafter \@firstoftwo
 \else \expandafter \@secondoftwo
 \fi
}%
\providecommand \@ifx [1]{%
 \ifx #1\expandafter \@firstoftwo
 \else \expandafter \@secondoftwo
 \fi
}%
\providecommand \natexlab [1]{#1}%
\providecommand \enquote  [1]{``#1''}%
\providecommand \bibnamefont  [1]{#1}%
\providecommand \bibfnamefont [1]{#1}%
\providecommand \citenamefont [1]{#1}%
\providecommand \href@noop [0]{\@secondoftwo}%
\providecommand \href [0]{\begingroup \@sanitize@url \@href}%
\providecommand \@href[1]{\@@startlink{#1}\@@href}%
\providecommand \@@href[1]{\endgroup#1\@@endlink}%
\providecommand \@sanitize@url [0]{\catcode `\\12\catcode `\$12\catcode
  `\&12\catcode `\#12\catcode `\^12\catcode `\_12\catcode `\%12\relax}%
\providecommand \@@startlink[1]{}%
\providecommand \@@endlink[0]{}%
\providecommand \url  [0]{\begingroup\@sanitize@url \@url }%
\providecommand \@url [1]{\endgroup\@href {#1}{\urlprefix }}%
\providecommand \urlprefix  [0]{URL }%
\providecommand \Eprint [0]{\href }%
\providecommand \doibase [0]{http://dx.doi.org/}%
\providecommand \selectlanguage [0]{\@gobble}%
\providecommand \bibinfo  [0]{\@secondoftwo}%
\providecommand \bibfield  [0]{\@secondoftwo}%
\providecommand \translation [1]{[#1]}%
\providecommand \BibitemOpen [0]{}%
\providecommand \bibitemStop [0]{}%
\providecommand \bibitemNoStop [0]{.\EOS\space}%
\providecommand \EOS [0]{\spacefactor3000\relax}%
\providecommand \BibitemShut  [1]{\csname bibitem#1\endcsname}%
\let\auto@bib@innerbib\@empty
\bibitem [{\citenamefont {Basov}\ \emph {et~al.}(2017)\citenamefont {Basov},
  \citenamefont {Averitt},\ and\ \citenamefont {Hsieh}}]{Basov:2017}%
  \BibitemOpen
  \bibfield  {author} {\bibinfo {author} {\bibfnamefont {D.~N.}\ \bibnamefont
  {Basov}}, \bibinfo {author} {\bibfnamefont {R.~D.}\ \bibnamefont {Averitt}},
  \ and\ \bibinfo {author} {\bibfnamefont {D.}~\bibnamefont {Hsieh}},\
  }\bibfield  {title} {\enquote {\bibinfo {title} {Towards properties on demand
  in quantum materials},}\ }\href {\doibase 10.1038/nmat5017} {\bibfield
  {journal} {\bibinfo  {journal} {Nature Materials}\ }\textbf {\bibinfo
  {volume} {16}},\ \bibinfo {pages} {1077--1088} (\bibinfo {year}
  {2017})}\BibitemShut {NoStop}%
\bibitem [{\citenamefont {Moser}\ \emph {et~al.}(2013)\citenamefont {Moser},
  \citenamefont {Moreschini}, \citenamefont {Ja\ifmmode \acute{c}\else
  \'{c}\fi{}imovi\ifmmode~\acute{c}\else \'{c}\fi{}}, \citenamefont
  {Bari\ifmmode \check{s}\else \v{s}\fi{}i\ifmmode~\acute{c}\else \'{c}\fi{}},
  \citenamefont {Berger}, \citenamefont {Magrez}, \citenamefont {Chang},
  \citenamefont {Kim}, \citenamefont {Bostwick}, \citenamefont {Rotenberg},
  \citenamefont {Forr\'o},\ and\ \citenamefont {Grioni}}]{Moser:2013}%
  \BibitemOpen
  \bibfield  {author} {\bibinfo {author} {\bibfnamefont {S.}~\bibnamefont
  {Moser}}, \bibinfo {author} {\bibfnamefont {L.}~\bibnamefont {Moreschini}},
  \bibinfo {author} {\bibfnamefont {J.}~\bibnamefont {Ja\ifmmode \acute{c}\else
  \'{c}\fi{}imovi\ifmmode~\acute{c}\else \'{c}\fi{}}}, \bibinfo {author}
  {\bibfnamefont {O.~S.}\ \bibnamefont {Bari\ifmmode \check{s}\else
  \v{s}\fi{}i\ifmmode~\acute{c}\else \'{c}\fi{}}}, \bibinfo {author}
  {\bibfnamefont {H.}~\bibnamefont {Berger}}, \bibinfo {author} {\bibfnamefont
  {A.}~\bibnamefont {Magrez}}, \bibinfo {author} {\bibfnamefont {Y.~J.}\
  \bibnamefont {Chang}}, \bibinfo {author} {\bibfnamefont {K.~S.}\ \bibnamefont
  {Kim}}, \bibinfo {author} {\bibfnamefont {A.}~\bibnamefont {Bostwick}},
  \bibinfo {author} {\bibfnamefont {E.}~\bibnamefont {Rotenberg}}, \bibinfo
  {author} {\bibfnamefont {L.}~\bibnamefont {Forr\'o}}, \ and\ \bibinfo
  {author} {\bibfnamefont {M.}~\bibnamefont {Grioni}},\ }\bibfield  {title}
  {\enquote {\bibinfo {title} {Tunable polaronic conduction in anatase
  {T}i{O}$_{2}$},}\ }\href {\doibase 10.1103/PhysRevLett.110.196403} {\bibfield
   {journal} {\bibinfo  {journal} {Phys. Rev. Lett.}\ }\textbf {\bibinfo
  {volume} {110}},\ \bibinfo {pages} {196403} (\bibinfo {year}
  {2013})}\BibitemShut {NoStop}%
\bibitem [{\citenamefont {Wang}\ \emph {et~al.}(2016)\citenamefont {Wang},
  \citenamefont {McKeown~Walker}, \citenamefont {Tamai}, \citenamefont {Wang},
  \citenamefont {Ristic}, \citenamefont {Bruno}, \citenamefont {de~la Torre},
  \citenamefont {Ricc{\`o}}, \citenamefont {Plumb}, \citenamefont {Shi},
  \citenamefont {Hlawenka}, \citenamefont {S{\'a}nchez-Barriga}, \citenamefont
  {Varykhalov}, \citenamefont {Kim}, \citenamefont {Hoesch}, \citenamefont
  {King}, \citenamefont {Meevasana}, \citenamefont {Diebold}, \citenamefont
  {Mesot}, \citenamefont {Moritz}, \citenamefont {Devereaux}, \citenamefont
  {Radovic},\ and\ \citenamefont {Baumberger}}]{McKeown:2016}%
  \BibitemOpen
  \bibfield  {author} {\bibinfo {author} {\bibfnamefont {Z.}~\bibnamefont
  {Wang}}, \bibinfo {author} {\bibfnamefont {S.}~\bibnamefont
  {McKeown~Walker}}, \bibinfo {author} {\bibfnamefont {A.}~\bibnamefont
  {Tamai}}, \bibinfo {author} {\bibfnamefont {Y.}~\bibnamefont {Wang}},
  \bibinfo {author} {\bibfnamefont {Z.}~\bibnamefont {Ristic}}, \bibinfo
  {author} {\bibfnamefont {F.~Y.}\ \bibnamefont {Bruno}}, \bibinfo {author}
  {\bibfnamefont {A.}~\bibnamefont {de~la Torre}}, \bibinfo {author}
  {\bibfnamefont {S.}~\bibnamefont {Ricc{\`o}}}, \bibinfo {author}
  {\bibfnamefont {N.~C.}\ \bibnamefont {Plumb}}, \bibinfo {author}
  {\bibfnamefont {M.}~\bibnamefont {Shi}}, \bibinfo {author} {\bibfnamefont
  {P.}~\bibnamefont {Hlawenka}}, \bibinfo {author} {\bibfnamefont
  {J.}~\bibnamefont {S{\'a}nchez-Barriga}}, \bibinfo {author} {\bibfnamefont
  {A.}~\bibnamefont {Varykhalov}}, \bibinfo {author} {\bibfnamefont {T.~K.}\
  \bibnamefont {Kim}}, \bibinfo {author} {\bibfnamefont {M.}~\bibnamefont
  {Hoesch}}, \bibinfo {author} {\bibfnamefont {P.~D.~C.}\ \bibnamefont {King}},
  \bibinfo {author} {\bibfnamefont {W.}~\bibnamefont {Meevasana}}, \bibinfo
  {author} {\bibfnamefont {U.}~\bibnamefont {Diebold}}, \bibinfo {author}
  {\bibfnamefont {J.}~\bibnamefont {Mesot}}, \bibinfo {author} {\bibfnamefont
  {B.}~\bibnamefont {Moritz}}, \bibinfo {author} {\bibfnamefont {T.~P.}\
  \bibnamefont {Devereaux}}, \bibinfo {author} {\bibfnamefont {M.}~\bibnamefont
  {Radovic}}, \ and\ \bibinfo {author} {\bibfnamefont {F.}~\bibnamefont
  {Baumberger}},\ }\bibfield  {title} {\enquote {\bibinfo {title} {Tailoring
  the nature and strength of electron--phonon interactions in the
  {Sr}{T}i{O}$_3$(001) 2{D} electron liquid},}\ }\href {\doibase
  10.1038/nmat4623} {\bibfield  {journal} {\bibinfo  {journal} {Nature
  Materials}\ }\textbf {\bibinfo {volume} {15}},\ \bibinfo {pages} {835--839}
  (\bibinfo {year} {2016})}\BibitemShut {NoStop}%
\bibitem [{\citenamefont {Kang}\ \emph {et~al.}(2018)\citenamefont {Kang},
  \citenamefont {Jung}, \citenamefont {Shin}, \citenamefont {Sohn},
  \citenamefont {Ryu}, \citenamefont {Kim}, \citenamefont {Hoesch},\ and\
  \citenamefont {Kim}}]{Kang:2018}%
  \BibitemOpen
  \bibfield  {author} {\bibinfo {author} {\bibfnamefont {Mingu}\ \bibnamefont
  {Kang}}, \bibinfo {author} {\bibfnamefont {Sung~Won}\ \bibnamefont {Jung}},
  \bibinfo {author} {\bibfnamefont {Woo~Jong}\ \bibnamefont {Shin}}, \bibinfo
  {author} {\bibfnamefont {Yeongsup}\ \bibnamefont {Sohn}}, \bibinfo {author}
  {\bibfnamefont {Sae~Hee}\ \bibnamefont {Ryu}}, \bibinfo {author}
  {\bibfnamefont {Timur~K.}\ \bibnamefont {Kim}}, \bibinfo {author}
  {\bibfnamefont {Moritz}\ \bibnamefont {Hoesch}}, \ and\ \bibinfo {author}
  {\bibfnamefont {Keun~Su}\ \bibnamefont {Kim}},\ }\bibfield  {title} {\enquote
  {\bibinfo {title} {Holstein polaron in a valley-degenerate two-dimensional
  semiconductor},}\ }\href {\doibase 10.1038/s41563-018-0092-7} {\bibfield
  {journal} {\bibinfo  {journal} {Nature Materials}\ }\textbf {\bibinfo
  {volume} {17}},\ \bibinfo {pages} {676--680} (\bibinfo {year}
  {2018})}\BibitemShut {NoStop}%
\bibitem [{\citenamefont {Chen}\ \emph {et~al.}(2018)\citenamefont {Chen},
  \citenamefont {Avila}, \citenamefont {Wang}, \citenamefont {Wang},
  \citenamefont {Mucha-Kruczy{\'n}ski}, \citenamefont {Shen}, \citenamefont
  {Yang}, \citenamefont {Nosarzewski}, \citenamefont {Devereaux}, \citenamefont
  {Zhang},\ and\ \citenamefont {Asensio}}]{ChenC:2018}%
  \BibitemOpen
  \bibfield  {author} {\bibinfo {author} {\bibfnamefont {Chaoyu}\ \bibnamefont
  {Chen}}, \bibinfo {author} {\bibfnamefont {Jos{\'e}}\ \bibnamefont {Avila}},
  \bibinfo {author} {\bibfnamefont {Shuopei}\ \bibnamefont {Wang}}, \bibinfo
  {author} {\bibfnamefont {Yao}\ \bibnamefont {Wang}}, \bibinfo {author}
  {\bibfnamefont {Marcin}\ \bibnamefont {Mucha-Kruczy{\'n}ski}}, \bibinfo
  {author} {\bibfnamefont {Cheng}\ \bibnamefont {Shen}}, \bibinfo {author}
  {\bibfnamefont {Rong}\ \bibnamefont {Yang}}, \bibinfo {author} {\bibfnamefont
  {Benjamin}\ \bibnamefont {Nosarzewski}}, \bibinfo {author} {\bibfnamefont
  {Thomas~P.}\ \bibnamefont {Devereaux}}, \bibinfo {author} {\bibfnamefont
  {Guangyu}\ \bibnamefont {Zhang}}, \ and\ \bibinfo {author} {\bibfnamefont
  {Maria~Carmen}\ \bibnamefont {Asensio}},\ }\bibfield  {title} {\enquote
  {\bibinfo {title} {Emergence of interfacial polarons from electron--phonon
  coupling in graphene/h-{BN} van der waals heterostructures},}\ }\href
  {\doibase 10.1021/acs.nanolett.7b04604} {\bibfield  {journal} {\bibinfo
  {journal} {Nano Letters}\ }\textbf {\bibinfo {volume} {18}},\ \bibinfo
  {pages} {1082--1087} (\bibinfo {year} {2018})}\BibitemShut {NoStop}%
\bibitem [{\citenamefont {Xiang}\ \emph {et~al.}(2023)\citenamefont {Xiang},
  \citenamefont {Ma}, \citenamefont {Gao}, \citenamefont {Guo}, \citenamefont
  {Huang}, \citenamefont {Xing}, \citenamefont {Tan}, \citenamefont {Zhao},
  \citenamefont {Wang},\ and\ \citenamefont {Shao}}]{Xiang:2023}%
  \BibitemOpen
  \bibfield  {author} {\bibinfo {author} {\bibfnamefont {Miaomiao}\
  \bibnamefont {Xiang}}, \bibinfo {author} {\bibfnamefont {Xiaochuan}\
  \bibnamefont {Ma}}, \bibinfo {author} {\bibfnamefont {Chang}\ \bibnamefont
  {Gao}}, \bibinfo {author} {\bibfnamefont {Ziyang}\ \bibnamefont {Guo}},
  \bibinfo {author} {\bibfnamefont {Chenxi}\ \bibnamefont {Huang}}, \bibinfo
  {author} {\bibfnamefont {Yue}\ \bibnamefont {Xing}}, \bibinfo {author}
  {\bibfnamefont {Shijing}\ \bibnamefont {Tan}}, \bibinfo {author}
  {\bibfnamefont {Jin}\ \bibnamefont {Zhao}}, \bibinfo {author} {\bibfnamefont
  {Bing}\ \bibnamefont {Wang}}, \ and\ \bibinfo {author} {\bibfnamefont
  {Xiang}\ \bibnamefont {Shao}},\ }\bibfield  {title} {\enquote {\bibinfo
  {title} {Revealing the polaron state at the {MoS}$_2$/{TiO}$_2$ interface},}\
  }\href {\doibase 10.1021/acs.jpclett.2c03856} {\bibfield  {journal} {\bibinfo
   {journal} {The Journal of Physical Chemistry Letters}\ }\textbf {\bibinfo
  {volume} {14}},\ \bibinfo {pages} {3360--3367} (\bibinfo {year}
  {2023})}\BibitemShut {NoStop}%
\bibitem [{\citenamefont {Caruso}\ \emph {et~al.}(2015)\citenamefont {Caruso},
  \citenamefont {Lambert},\ and\ \citenamefont {Giustino}}]{Caruso:2015}%
  \BibitemOpen
  \bibfield  {author} {\bibinfo {author} {\bibfnamefont {Fabio}\ \bibnamefont
  {Caruso}}, \bibinfo {author} {\bibfnamefont {Henry}\ \bibnamefont {Lambert}},
  \ and\ \bibinfo {author} {\bibfnamefont {Feliciano}\ \bibnamefont
  {Giustino}},\ }\bibfield  {title} {\enquote {\bibinfo {title} {Band
  structures of plasmonic polarons},}\ }\href {\doibase
  10.1103/PhysRevLett.114.146404} {\bibfield  {journal} {\bibinfo  {journal}
  {Phys. Rev. Lett.}\ }\textbf {\bibinfo {volume} {114}},\ \bibinfo {pages}
  {146404} (\bibinfo {year} {2015})}\BibitemShut {NoStop}%
\bibitem [{\citenamefont {Riley}\ \emph {et~al.}(2018)\citenamefont {Riley},
  \citenamefont {Caruso}, \citenamefont {Verdi}, \citenamefont {Duffy},
  \citenamefont {Watson}, \citenamefont {Bawden}, \citenamefont {Volckaert},
  \citenamefont {van~der Laan}, \citenamefont {Hesjedal}, \citenamefont
  {Hoesch}, \citenamefont {Giustino},\ and\ \citenamefont {King}}]{Riley:2018}%
  \BibitemOpen
  \bibfield  {author} {\bibinfo {author} {\bibfnamefont {J.~M.}\ \bibnamefont
  {Riley}}, \bibinfo {author} {\bibfnamefont {F.}~\bibnamefont {Caruso}},
  \bibinfo {author} {\bibfnamefont {C.}~\bibnamefont {Verdi}}, \bibinfo
  {author} {\bibfnamefont {L.~B.}\ \bibnamefont {Duffy}}, \bibinfo {author}
  {\bibfnamefont {M.~D.}\ \bibnamefont {Watson}}, \bibinfo {author}
  {\bibfnamefont {L.}~\bibnamefont {Bawden}}, \bibinfo {author} {\bibfnamefont
  {K.}~\bibnamefont {Volckaert}}, \bibinfo {author} {\bibfnamefont
  {G.}~\bibnamefont {van~der Laan}}, \bibinfo {author} {\bibfnamefont
  {T.}~\bibnamefont {Hesjedal}}, \bibinfo {author} {\bibfnamefont
  {M.}~\bibnamefont {Hoesch}}, \bibinfo {author} {\bibfnamefont
  {F.}~\bibnamefont {Giustino}}, \ and\ \bibinfo {author} {\bibfnamefont
  {P.~D.~C.}\ \bibnamefont {King}},\ }\bibfield  {title} {\enquote {\bibinfo
  {title} {Crossover from lattice to plasmonic polarons of a spin-polarised
  electron gas in ferromagnetic {EuO}},}\ }\href {\doibase
  10.1038/s41467-018-04749-w} {\bibfield  {journal} {\bibinfo  {journal}
  {Nature Communications}\ }\textbf {\bibinfo {volume} {9}},\ \bibinfo {pages}
  {2305} (\bibinfo {year} {2018})}\BibitemShut {NoStop}%
\bibitem [{\citenamefont {Ma}\ \emph {et~al.}(2021)\citenamefont {Ma},
  \citenamefont {Cheng}, \citenamefont {Tian}, \citenamefont {Liu},
  \citenamefont {Cui}, \citenamefont {Huang}, \citenamefont {Tan},
  \citenamefont {Yang},\ and\ \citenamefont {Wang}}]{Xiaochuan:2021}%
  \BibitemOpen
  \bibfield  {author} {\bibinfo {author} {\bibfnamefont {Xiaochuan}\
  \bibnamefont {Ma}}, \bibinfo {author} {\bibfnamefont {Zhengwang}\
  \bibnamefont {Cheng}}, \bibinfo {author} {\bibfnamefont {Mingyang}\
  \bibnamefont {Tian}}, \bibinfo {author} {\bibfnamefont {Xiaofeng}\
  \bibnamefont {Liu}}, \bibinfo {author} {\bibfnamefont {Xuefeng}\ \bibnamefont
  {Cui}}, \bibinfo {author} {\bibfnamefont {Yaobo}\ \bibnamefont {Huang}},
  \bibinfo {author} {\bibfnamefont {Shijing}\ \bibnamefont {Tan}}, \bibinfo
  {author} {\bibfnamefont {Jinlong}\ \bibnamefont {Yang}}, \ and\ \bibinfo
  {author} {\bibfnamefont {Bing}\ \bibnamefont {Wang}},\ }\bibfield  {title}
  {\enquote {\bibinfo {title} {Formation of plasmonic polarons in highly
  electron-doped anatase {TiO}$_2$},}\ }\href {\doibase
  10.1021/acs.nanolett.0c03802} {\bibfield  {journal} {\bibinfo  {journal}
  {Nano Letters}\ }\textbf {\bibinfo {volume} {21}},\ \bibinfo {pages}
  {430--436} (\bibinfo {year} {2021})}\BibitemShut {NoStop}%
\bibitem [{\citenamefont {Caruso}\ \emph {et~al.}(2021)\citenamefont {Caruso},
  \citenamefont {Amsalem}, \citenamefont {Ma}, \citenamefont {Aljarb},
  \citenamefont {Schultz}, \citenamefont {Zacharias}, \citenamefont {Tung},
  \citenamefont {Koch},\ and\ \citenamefont {Draxl}}]{Caruso:2021}%
  \BibitemOpen
  \bibfield  {author} {\bibinfo {author} {\bibfnamefont {Fabio}\ \bibnamefont
  {Caruso}}, \bibinfo {author} {\bibfnamefont {Patrick}\ \bibnamefont
  {Amsalem}}, \bibinfo {author} {\bibfnamefont {Jie}\ \bibnamefont {Ma}},
  \bibinfo {author} {\bibfnamefont {Areej}\ \bibnamefont {Aljarb}}, \bibinfo
  {author} {\bibfnamefont {Thorsten}\ \bibnamefont {Schultz}}, \bibinfo
  {author} {\bibfnamefont {Marios}\ \bibnamefont {Zacharias}}, \bibinfo
  {author} {\bibfnamefont {Vincent}\ \bibnamefont {Tung}}, \bibinfo {author}
  {\bibfnamefont {Norbert}\ \bibnamefont {Koch}}, \ and\ \bibinfo {author}
  {\bibfnamefont {Claudia}\ \bibnamefont {Draxl}},\ }\bibfield  {title}
  {\enquote {\bibinfo {title} {Two-dimensional plasmonic polarons in $n$-doped
  monolayer {MoS}$_2$},}\ }\href {\doibase 10.1103/PhysRevB.103.205152}
  {\bibfield  {journal} {\bibinfo  {journal} {Phys. Rev. B}\ }\textbf {\bibinfo
  {volume} {103}},\ \bibinfo {pages} {205152} (\bibinfo {year}
  {2021})}\BibitemShut {NoStop}%
\bibitem [{\citenamefont {Britnell}\ \emph {et~al.}(2013)\citenamefont
  {Britnell}, \citenamefont {Ribeiro}, \citenamefont {Eckmann}, \citenamefont
  {Jalil}, \citenamefont {Belle}, \citenamefont {Mishchenko}, \citenamefont
  {Kim}, \citenamefont {Gorbachev}, \citenamefont {Georgiou}, \citenamefont
  {Morozov}, \citenamefont {Grigorenko}, \citenamefont {Geim}, \citenamefont
  {Casiraghi}, \citenamefont {Neto},\ and\ \citenamefont
  {Novoselov}}]{Britnell:2013}%
  \BibitemOpen
  \bibfield  {author} {\bibinfo {author} {\bibfnamefont {L.}~\bibnamefont
  {Britnell}}, \bibinfo {author} {\bibfnamefont {R.~M.}\ \bibnamefont
  {Ribeiro}}, \bibinfo {author} {\bibfnamefont {A.}~\bibnamefont {Eckmann}},
  \bibinfo {author} {\bibfnamefont {R.}~\bibnamefont {Jalil}}, \bibinfo
  {author} {\bibfnamefont {B.~D.}\ \bibnamefont {Belle}}, \bibinfo {author}
  {\bibfnamefont {A.}~\bibnamefont {Mishchenko}}, \bibinfo {author}
  {\bibfnamefont {Y.-J.}\ \bibnamefont {Kim}}, \bibinfo {author} {\bibfnamefont
  {R.~V.}\ \bibnamefont {Gorbachev}}, \bibinfo {author} {\bibfnamefont
  {T.}~\bibnamefont {Georgiou}}, \bibinfo {author} {\bibfnamefont {S.~V.}\
  \bibnamefont {Morozov}}, \bibinfo {author} {\bibfnamefont {A.~N.}\
  \bibnamefont {Grigorenko}}, \bibinfo {author} {\bibfnamefont {A.~K.}\
  \bibnamefont {Geim}}, \bibinfo {author} {\bibfnamefont {C.}~\bibnamefont
  {Casiraghi}}, \bibinfo {author} {\bibfnamefont {A.~H.~Castro}\ \bibnamefont
  {Neto}}, \ and\ \bibinfo {author} {\bibfnamefont {K.~S.}\ \bibnamefont
  {Novoselov}},\ }\bibfield  {title} {\enquote {\bibinfo {title} {Strong
  light-matter interactions in heterostructures of atomically thin films},}\
  }\href {\doibase 10.1126/science.1235547} {\bibfield  {journal} {\bibinfo
  {journal} {Science}\ }\textbf {\bibinfo {volume} {340}},\ \bibinfo {pages}
  {1311--1314} (\bibinfo {year} {2013})}\BibitemShut {NoStop}%
\bibitem [{\citenamefont {Zihlmann}\ \emph {et~al.}(2018)\citenamefont
  {Zihlmann}, \citenamefont {Cummings}, \citenamefont {Garcia}, \citenamefont
  {Kedves}, \citenamefont {Watanabe}, \citenamefont {Taniguchi}, \citenamefont
  {Sch\"onenberger},\ and\ \citenamefont {Makk}}]{Zihlmann:2018}%
  \BibitemOpen
  \bibfield  {author} {\bibinfo {author} {\bibfnamefont {Simon}\ \bibnamefont
  {Zihlmann}}, \bibinfo {author} {\bibfnamefont {Aron~W.}\ \bibnamefont
  {Cummings}}, \bibinfo {author} {\bibfnamefont {Jose~H.}\ \bibnamefont
  {Garcia}}, \bibinfo {author} {\bibfnamefont {M\'at\'e}\ \bibnamefont
  {Kedves}}, \bibinfo {author} {\bibfnamefont {Kenji}\ \bibnamefont
  {Watanabe}}, \bibinfo {author} {\bibfnamefont {Takashi}\ \bibnamefont
  {Taniguchi}}, \bibinfo {author} {\bibfnamefont {Christian}\ \bibnamefont
  {Sch\"onenberger}}, \ and\ \bibinfo {author} {\bibfnamefont {P\'eter}\
  \bibnamefont {Makk}},\ }\bibfield  {title} {\enquote {\bibinfo {title} {Large
  spin relaxation anisotropy and valley-zeeman spin-orbit coupling in
  {WSe}$_2$/graphene/$h$-{BN} heterostructures},}\ }\href {\doibase
  10.1103/PhysRevB.97.075434} {\bibfield  {journal} {\bibinfo  {journal} {Phys.
  Rev. B}\ }\textbf {\bibinfo {volume} {97}},\ \bibinfo {pages} {075434}
  (\bibinfo {year} {2018})}\BibitemShut {NoStop}%
\bibitem [{\citenamefont {Arora}\ \emph {et~al.}(2020)\citenamefont {Arora},
  \citenamefont {Polski}, \citenamefont {Zhang}, \citenamefont {Thomson},
  \citenamefont {Choi}, \citenamefont {Kim}, \citenamefont {Lin}, \citenamefont
  {Wilson}, \citenamefont {Xu}, \citenamefont {Chu}, \citenamefont {Watanabe},
  \citenamefont {Taniguchi}, \citenamefont {Alicea},\ and\ \citenamefont
  {Nadj-Perge}}]{Arora:2020}%
  \BibitemOpen
  \bibfield  {author} {\bibinfo {author} {\bibfnamefont {Harpreet~Singh}\
  \bibnamefont {Arora}}, \bibinfo {author} {\bibfnamefont {Robert}\
  \bibnamefont {Polski}}, \bibinfo {author} {\bibfnamefont {Yiran}\
  \bibnamefont {Zhang}}, \bibinfo {author} {\bibfnamefont {Alex}\ \bibnamefont
  {Thomson}}, \bibinfo {author} {\bibfnamefont {Youngjoon}\ \bibnamefont
  {Choi}}, \bibinfo {author} {\bibfnamefont {Hyunjin}\ \bibnamefont {Kim}},
  \bibinfo {author} {\bibfnamefont {Zhong}\ \bibnamefont {Lin}}, \bibinfo
  {author} {\bibfnamefont {Ilham~Zaky}\ \bibnamefont {Wilson}}, \bibinfo
  {author} {\bibfnamefont {Xiaodong}\ \bibnamefont {Xu}}, \bibinfo {author}
  {\bibfnamefont {Jiun-Haw}\ \bibnamefont {Chu}}, \bibinfo {author}
  {\bibfnamefont {Kenji}\ \bibnamefont {Watanabe}}, \bibinfo {author}
  {\bibfnamefont {Takashi}\ \bibnamefont {Taniguchi}}, \bibinfo {author}
  {\bibfnamefont {Jason}\ \bibnamefont {Alicea}}, \ and\ \bibinfo {author}
  {\bibfnamefont {Stevan}\ \bibnamefont {Nadj-Perge}},\ }\bibfield  {title}
  {\enquote {\bibinfo {title} {Superconductivity in metallic twisted bilayer
  graphene stabilized by {WSe}$_2$},}\ }\href {\doibase
  10.1038/s41586-020-2473-8} {\bibfield  {journal} {\bibinfo  {journal}
  {Nature}\ }\textbf {\bibinfo {volume} {583}},\ \bibinfo {pages} {379--384}
  (\bibinfo {year} {2020})}\BibitemShut {NoStop}%
\bibitem [{\citenamefont {Tang}\ \emph {et~al.}(2020)\citenamefont {Tang},
  \citenamefont {Li}, \citenamefont {Li}, \citenamefont {Xu}, \citenamefont
  {Liu}, \citenamefont {Barmak}, \citenamefont {Watanabe}, \citenamefont
  {Taniguchi}, \citenamefont {MacDonald}, \citenamefont {Shan},\ and\
  \citenamefont {Mak}}]{Tang:2020}%
  \BibitemOpen
  \bibfield  {author} {\bibinfo {author} {\bibfnamefont {Yanhao}\ \bibnamefont
  {Tang}}, \bibinfo {author} {\bibfnamefont {Lizhong}\ \bibnamefont {Li}},
  \bibinfo {author} {\bibfnamefont {Tingxin}\ \bibnamefont {Li}}, \bibinfo
  {author} {\bibfnamefont {Yang}\ \bibnamefont {Xu}}, \bibinfo {author}
  {\bibfnamefont {Song}\ \bibnamefont {Liu}}, \bibinfo {author} {\bibfnamefont
  {Katayun}\ \bibnamefont {Barmak}}, \bibinfo {author} {\bibfnamefont {Kenji}\
  \bibnamefont {Watanabe}}, \bibinfo {author} {\bibfnamefont {Takashi}\
  \bibnamefont {Taniguchi}}, \bibinfo {author} {\bibfnamefont {Allan~H.}\
  \bibnamefont {MacDonald}}, \bibinfo {author} {\bibfnamefont {Jie}\
  \bibnamefont {Shan}}, \ and\ \bibinfo {author} {\bibfnamefont {Kin~Fai}\
  \bibnamefont {Mak}},\ }\bibfield  {title} {\enquote {\bibinfo {title}
  {Simulation of hubbard model physics in {WSe}$_2$/{WS}$_2$ moir{\'e}
  superlattices},}\ }\href {\doibase 10.1038/s41586-020-2085-3} {\bibfield
  {journal} {\bibinfo  {journal} {Nature}\ }\textbf {\bibinfo {volume} {579}},\
  \bibinfo {pages} {353--358} (\bibinfo {year} {2020})}\BibitemShut {NoStop}%
\bibitem [{\citenamefont {Regan}\ \emph {et~al.}(2020)\citenamefont {Regan},
  \citenamefont {Wang}, \citenamefont {Jin}, \citenamefont {Bakti~Utama},
  \citenamefont {Gao}, \citenamefont {Wei}, \citenamefont {Zhao}, \citenamefont
  {Zhao}, \citenamefont {Zhang}, \citenamefont {Yumigeta}, \citenamefont
  {Blei}, \citenamefont {Carlstr{\"o}m}, \citenamefont {Watanabe},
  \citenamefont {Taniguchi}, \citenamefont {Tongay}, \citenamefont {Crommie},
  \citenamefont {Zettl},\ and\ \citenamefont {Wang}}]{Regan:2020}%
  \BibitemOpen
  \bibfield  {author} {\bibinfo {author} {\bibfnamefont {Emma~C.}\ \bibnamefont
  {Regan}}, \bibinfo {author} {\bibfnamefont {Danqing}\ \bibnamefont {Wang}},
  \bibinfo {author} {\bibfnamefont {Chenhao}\ \bibnamefont {Jin}}, \bibinfo
  {author} {\bibfnamefont {M.~Iqbal}\ \bibnamefont {Bakti~Utama}}, \bibinfo
  {author} {\bibfnamefont {Beini}\ \bibnamefont {Gao}}, \bibinfo {author}
  {\bibfnamefont {Xin}\ \bibnamefont {Wei}}, \bibinfo {author} {\bibfnamefont
  {Sihan}\ \bibnamefont {Zhao}}, \bibinfo {author} {\bibfnamefont {Wenyu}\
  \bibnamefont {Zhao}}, \bibinfo {author} {\bibfnamefont {Zuocheng}\
  \bibnamefont {Zhang}}, \bibinfo {author} {\bibfnamefont {Kentaro}\
  \bibnamefont {Yumigeta}}, \bibinfo {author} {\bibfnamefont {Mark}\
  \bibnamefont {Blei}}, \bibinfo {author} {\bibfnamefont {Johan~D.}\
  \bibnamefont {Carlstr{\"o}m}}, \bibinfo {author} {\bibfnamefont {Kenji}\
  \bibnamefont {Watanabe}}, \bibinfo {author} {\bibfnamefont {Takashi}\
  \bibnamefont {Taniguchi}}, \bibinfo {author} {\bibfnamefont {Sefaattin}\
  \bibnamefont {Tongay}}, \bibinfo {author} {\bibfnamefont {Michael}\
  \bibnamefont {Crommie}}, \bibinfo {author} {\bibfnamefont {Alex}\
  \bibnamefont {Zettl}}, \ and\ \bibinfo {author} {\bibfnamefont {Feng}\
  \bibnamefont {Wang}},\ }\bibfield  {title} {\enquote {\bibinfo {title} {Mott
  and generalized wigner crystal states in {WSe}$_2$/{WS}$_2$ moir{\'e}
  superlattices},}\ }\href {\doibase 10.1038/s41586-020-2092-4} {\bibfield
  {journal} {\bibinfo  {journal} {Nature}\ }\textbf {\bibinfo {volume} {579}},\
  \bibinfo {pages} {359--363} (\bibinfo {year} {2020})}\BibitemShut {NoStop}%
\bibitem [{\citenamefont {Kennes}\ \emph {et~al.}(2021)\citenamefont {Kennes},
  \citenamefont {Claassen}, \citenamefont {Xian}, \citenamefont {Georges},
  \citenamefont {Millis}, \citenamefont {Hone}, \citenamefont {Dean},
  \citenamefont {Basov}, \citenamefont {Pasupathy},\ and\ \citenamefont
  {Rubio}}]{Kennes:2021}%
  \BibitemOpen
  \bibfield  {author} {\bibinfo {author} {\bibfnamefont {Dante~M.}\
  \bibnamefont {Kennes}}, \bibinfo {author} {\bibfnamefont {Martin}\
  \bibnamefont {Claassen}}, \bibinfo {author} {\bibfnamefont {Lede}\
  \bibnamefont {Xian}}, \bibinfo {author} {\bibfnamefont {Antoine}\
  \bibnamefont {Georges}}, \bibinfo {author} {\bibfnamefont {Andrew~J.}\
  \bibnamefont {Millis}}, \bibinfo {author} {\bibfnamefont {James}\
  \bibnamefont {Hone}}, \bibinfo {author} {\bibfnamefont {Cory~R.}\
  \bibnamefont {Dean}}, \bibinfo {author} {\bibfnamefont {D.~N.}\ \bibnamefont
  {Basov}}, \bibinfo {author} {\bibfnamefont {Abhay~N.}\ \bibnamefont
  {Pasupathy}}, \ and\ \bibinfo {author} {\bibfnamefont {Angel}\ \bibnamefont
  {Rubio}},\ }\bibfield  {title} {\enquote {\bibinfo {title} {Moir{\'e}
  heterostructures as a condensed-matter quantum simulator},}\ }\href {\doibase
  10.1038/s41567-020-01154-3} {\bibfield  {journal} {\bibinfo  {journal}
  {Nature Physics}\ }\textbf {\bibinfo {volume} {17}},\ \bibinfo {pages}
  {155--163} (\bibinfo {year} {2021})}\BibitemShut {NoStop}%
\bibitem [{\citenamefont {Ulstrup}\ \emph {et~al.}(2019)\citenamefont
  {Ulstrup}, \citenamefont {Giusca}, \citenamefont {Miwa}, \citenamefont
  {Sanders}, \citenamefont {Browning}, \citenamefont {Dudin}, \citenamefont
  {Cacho}, \citenamefont {Kazakova}, \citenamefont {Gaskill}, \citenamefont
  {Myers-Ward}, \citenamefont {Zhang}, \citenamefont {Terrones},\ and\
  \citenamefont {Hofmann}}]{Ulstrup2:2019}%
  \BibitemOpen
  \bibfield  {author} {\bibinfo {author} {\bibfnamefont {S{\o}ren}\
  \bibnamefont {Ulstrup}}, \bibinfo {author} {\bibfnamefont {Cristina~E.}\
  \bibnamefont {Giusca}}, \bibinfo {author} {\bibfnamefont {Jill~A.}\
  \bibnamefont {Miwa}}, \bibinfo {author} {\bibfnamefont {Charlotte~E.}\
  \bibnamefont {Sanders}}, \bibinfo {author} {\bibfnamefont {Alex}\
  \bibnamefont {Browning}}, \bibinfo {author} {\bibfnamefont {Pavel}\
  \bibnamefont {Dudin}}, \bibinfo {author} {\bibfnamefont {Cephise}\
  \bibnamefont {Cacho}}, \bibinfo {author} {\bibfnamefont {Olga}\ \bibnamefont
  {Kazakova}}, \bibinfo {author} {\bibfnamefont {D.~Kurt}\ \bibnamefont
  {Gaskill}}, \bibinfo {author} {\bibfnamefont {Rachael~L.}\ \bibnamefont
  {Myers-Ward}}, \bibinfo {author} {\bibfnamefont {Tianyi}\ \bibnamefont
  {Zhang}}, \bibinfo {author} {\bibfnamefont {Mauricio}\ \bibnamefont
  {Terrones}}, \ and\ \bibinfo {author} {\bibfnamefont {Philip}\ \bibnamefont
  {Hofmann}},\ }\bibfield  {title} {\enquote {\bibinfo {title} {Nanoscale
  mapping of quasiparticle band alignment},}\ }\href {\doibase
  10.1038/s41467-019-11253-2} {\bibfield  {journal} {\bibinfo  {journal}
  {Nature Communications}\ }\textbf {\bibinfo {volume} {10}},\ \bibinfo {pages}
  {3283} (\bibinfo {year} {2019})}\BibitemShut {NoStop}%
\bibitem [{\citenamefont {Waldecker}\ \emph {et~al.}(2019)\citenamefont
  {Waldecker}, \citenamefont {Raja}, \citenamefont {R\"osner}, \citenamefont
  {Steinke}, \citenamefont {Bostwick}, \citenamefont {Koch}, \citenamefont
  {Jozwiak}, \citenamefont {Taniguchi}, \citenamefont {Watanabe}, \citenamefont
  {Rotenberg}, \citenamefont {Wehling},\ and\ \citenamefont
  {Heinz}}]{Waldecker:2019}%
  \BibitemOpen
  \bibfield  {author} {\bibinfo {author} {\bibfnamefont {Lutz}\ \bibnamefont
  {Waldecker}}, \bibinfo {author} {\bibfnamefont {Archana}\ \bibnamefont
  {Raja}}, \bibinfo {author} {\bibfnamefont {Malte}\ \bibnamefont {R\"osner}},
  \bibinfo {author} {\bibfnamefont {Christina}\ \bibnamefont {Steinke}},
  \bibinfo {author} {\bibfnamefont {Aaron}\ \bibnamefont {Bostwick}}, \bibinfo
  {author} {\bibfnamefont {Roland~J.}\ \bibnamefont {Koch}}, \bibinfo {author}
  {\bibfnamefont {Chris}\ \bibnamefont {Jozwiak}}, \bibinfo {author}
  {\bibfnamefont {Takashi}\ \bibnamefont {Taniguchi}}, \bibinfo {author}
  {\bibfnamefont {Kenji}\ \bibnamefont {Watanabe}}, \bibinfo {author}
  {\bibfnamefont {Eli}\ \bibnamefont {Rotenberg}}, \bibinfo {author}
  {\bibfnamefont {Tim~O.}\ \bibnamefont {Wehling}}, \ and\ \bibinfo {author}
  {\bibfnamefont {Tony~F.}\ \bibnamefont {Heinz}},\ }\bibfield  {title}
  {\enquote {\bibinfo {title} {Rigid band shifts in two-dimensional
  semiconductors through external dielectric screening},}\ }\href {\doibase
  10.1103/PhysRevLett.123.206403} {\bibfield  {journal} {\bibinfo  {journal}
  {Phys. Rev. Lett.}\ }\textbf {\bibinfo {volume} {123}},\ \bibinfo {pages}
  {206403} (\bibinfo {year} {2019})}\BibitemShut {NoStop}%
\bibitem [{\citenamefont {Ulstrup}\ \emph {et~al.}(2020)\citenamefont
  {Ulstrup}, \citenamefont {Koch}, \citenamefont {Singh}, \citenamefont
  {McCreary}, \citenamefont {Jonker}, \citenamefont {Robinson}, \citenamefont
  {Jozwiak}, \citenamefont {Rotenberg}, \citenamefont {Bostwick}, \citenamefont
  {Katoch},\ and\ \citenamefont {Miwa}}]{Ulstrup:2020}%
  \BibitemOpen
  \bibfield  {author} {\bibinfo {author} {\bibfnamefont {Søren}\ \bibnamefont
  {Ulstrup}}, \bibinfo {author} {\bibfnamefont {Roland~J.}\ \bibnamefont
  {Koch}}, \bibinfo {author} {\bibfnamefont {Simranjeet}\ \bibnamefont
  {Singh}}, \bibinfo {author} {\bibfnamefont {Kathleen~M.}\ \bibnamefont
  {McCreary}}, \bibinfo {author} {\bibfnamefont {Berend~T.}\ \bibnamefont
  {Jonker}}, \bibinfo {author} {\bibfnamefont {Jeremy~T.}\ \bibnamefont
  {Robinson}}, \bibinfo {author} {\bibfnamefont {Chris}\ \bibnamefont
  {Jozwiak}}, \bibinfo {author} {\bibfnamefont {Eli}\ \bibnamefont
  {Rotenberg}}, \bibinfo {author} {\bibfnamefont {Aaron}\ \bibnamefont
  {Bostwick}}, \bibinfo {author} {\bibfnamefont {Jyoti}\ \bibnamefont
  {Katoch}}, \ and\ \bibinfo {author} {\bibfnamefont {Jill~A.}\ \bibnamefont
  {Miwa}},\ }\bibfield  {title} {\enquote {\bibinfo {title} {Direct observation
  of minibands in a twisted graphene/{WS}$_2$ bilayer},}\ }\href {\doibase
  10.1126/sciadv.aay6104} {\bibfield  {journal} {\bibinfo  {journal} {Science
  Advances}\ }\textbf {\bibinfo {volume} {6}},\ \bibinfo {pages} {eaay6104}
  (\bibinfo {year} {2020})}\BibitemShut {NoStop}%
\bibitem [{\citenamefont {Xie}\ \emph {et~al.}(2022)\citenamefont {Xie},
  \citenamefont {Faeth}, \citenamefont {Tang}, \citenamefont {Li},
  \citenamefont {Gerber}, \citenamefont {Parzyck}, \citenamefont {Chowdhury},
  \citenamefont {Zhang}, \citenamefont {Jozwiak}, \citenamefont {Bostwick},
  \citenamefont {Rotenberg}, \citenamefont {Kim}, \citenamefont {Shan},
  \citenamefont {Mak},\ and\ \citenamefont {Shen}}]{Xie:2022}%
  \BibitemOpen
  \bibfield  {author} {\bibinfo {author} {\bibfnamefont {Saien}\ \bibnamefont
  {Xie}}, \bibinfo {author} {\bibfnamefont {Brendan~D.}\ \bibnamefont {Faeth}},
  \bibinfo {author} {\bibfnamefont {Yanhao}\ \bibnamefont {Tang}}, \bibinfo
  {author} {\bibfnamefont {Lizhong}\ \bibnamefont {Li}}, \bibinfo {author}
  {\bibfnamefont {Eli}\ \bibnamefont {Gerber}}, \bibinfo {author}
  {\bibfnamefont {Christopher~T.}\ \bibnamefont {Parzyck}}, \bibinfo {author}
  {\bibfnamefont {Debanjan}\ \bibnamefont {Chowdhury}}, \bibinfo {author}
  {\bibfnamefont {Ya-Hui}\ \bibnamefont {Zhang}}, \bibinfo {author}
  {\bibfnamefont {Christopher}\ \bibnamefont {Jozwiak}}, \bibinfo {author}
  {\bibfnamefont {Aaron}\ \bibnamefont {Bostwick}}, \bibinfo {author}
  {\bibfnamefont {Eli}\ \bibnamefont {Rotenberg}}, \bibinfo {author}
  {\bibfnamefont {Eun-Ah}\ \bibnamefont {Kim}}, \bibinfo {author}
  {\bibfnamefont {Jie}\ \bibnamefont {Shan}}, \bibinfo {author} {\bibfnamefont
  {Kin~Fai}\ \bibnamefont {Mak}}, \ and\ \bibinfo {author} {\bibfnamefont
  {Kyle~M.}\ \bibnamefont {Shen}},\ }\bibfield  {title} {\enquote {\bibinfo
  {title} {Strong interlayer interactions in bilayer and trilayer moir{\'e}
  superlattices},}\ }\href {\doibase 10.1126/sciadv.abk1911} {\bibfield
  {journal} {\bibinfo  {journal} {Science Advances}\ }\textbf {\bibinfo
  {volume} {8}},\ \bibinfo {pages} {eabk1911} (\bibinfo {year}
  {2022})}\BibitemShut {NoStop}%
\bibitem [{\citenamefont {Hennighausen}\ \emph {et~al.}(2023)\citenamefont
  {Hennighausen}, \citenamefont {Moon}, \citenamefont {McCreary}, \citenamefont
  {Li}, \citenamefont {van~'t Erve},\ and\ \citenamefont
  {Jonker}}]{Hennighausen:2023}%
  \BibitemOpen
  \bibfield  {author} {\bibinfo {author} {\bibfnamefont {Zachariah}\
  \bibnamefont {Hennighausen}}, \bibinfo {author} {\bibfnamefont {Jisoo}\
  \bibnamefont {Moon}}, \bibinfo {author} {\bibfnamefont {Kathleen~M.}\
  \bibnamefont {McCreary}}, \bibinfo {author} {\bibfnamefont {Connie~H.}\
  \bibnamefont {Li}}, \bibinfo {author} {\bibfnamefont {Olaf M.~J.}\
  \bibnamefont {van~'t Erve}}, \ and\ \bibinfo {author} {\bibfnamefont
  {Berend~T.}\ \bibnamefont {Jonker}},\ }\bibfield  {title} {\enquote {\bibinfo
  {title} {Interlayer exciton--phonon bound state in {Bi}$_2${Se}$_3$/monolayer
  {WS}$_2$ van der waals heterostructures},}\ }\href {\doibase
  10.1021/acsnano.2c10313} {\bibfield  {journal} {\bibinfo  {journal} {ACS
  Nano}\ }\textbf {\bibinfo {volume} {17}},\ \bibinfo {pages} {2529--2536}
  (\bibinfo {year} {2023})}\BibitemShut {NoStop}%
\bibitem [{\citenamefont {Nguyen}\ \emph {et~al.}(2019)\citenamefont {Nguyen},
  \citenamefont {Teutsch}, \citenamefont {Wilson}, \citenamefont {Kahn},
  \citenamefont {Xia}, \citenamefont {Graham}, \citenamefont {Kandyba},
  \citenamefont {Giampietri}, \citenamefont {Barinov}, \citenamefont
  {Constantinescu}, \citenamefont {Yeung}, \citenamefont {Hine}, \citenamefont
  {Xu}, \citenamefont {Cobden},\ and\ \citenamefont {Wilson}}]{Nguyen:2019}%
  \BibitemOpen
  \bibfield  {author} {\bibinfo {author} {\bibfnamefont {Paul~V.}\ \bibnamefont
  {Nguyen}}, \bibinfo {author} {\bibfnamefont {Natalie~C.}\ \bibnamefont
  {Teutsch}}, \bibinfo {author} {\bibfnamefont {Nathan~P.}\ \bibnamefont
  {Wilson}}, \bibinfo {author} {\bibfnamefont {Joshua}\ \bibnamefont {Kahn}},
  \bibinfo {author} {\bibfnamefont {Xue}\ \bibnamefont {Xia}}, \bibinfo
  {author} {\bibfnamefont {Abigail~J.}\ \bibnamefont {Graham}}, \bibinfo
  {author} {\bibfnamefont {Viktor}\ \bibnamefont {Kandyba}}, \bibinfo {author}
  {\bibfnamefont {Alessio}\ \bibnamefont {Giampietri}}, \bibinfo {author}
  {\bibfnamefont {Alexei}\ \bibnamefont {Barinov}}, \bibinfo {author}
  {\bibfnamefont {Gabriel~C.}\ \bibnamefont {Constantinescu}}, \bibinfo
  {author} {\bibfnamefont {Nelson}\ \bibnamefont {Yeung}}, \bibinfo {author}
  {\bibfnamefont {Nicholas D.~M.}\ \bibnamefont {Hine}}, \bibinfo {author}
  {\bibfnamefont {Xiaodong}\ \bibnamefont {Xu}}, \bibinfo {author}
  {\bibfnamefont {David~H.}\ \bibnamefont {Cobden}}, \ and\ \bibinfo {author}
  {\bibfnamefont {Neil~R.}\ \bibnamefont {Wilson}},\ }\bibfield  {title}
  {\enquote {\bibinfo {title} {Visualizing electrostatic gating effects in
  two-dimensional heterostructures},}\ }\href {\doibase
  10.1038/s41586-019-1402-1} {\bibfield  {journal} {\bibinfo  {journal}
  {Nature}\ }\textbf {\bibinfo {volume} {572}},\ \bibinfo {pages} {220--223}
  (\bibinfo {year} {2019})}\BibitemShut {NoStop}%
\bibitem [{\citenamefont {Chuang}\ \emph {et~al.}(2014)\citenamefont {Chuang},
  \citenamefont {Tan}, \citenamefont {Ghimire}, \citenamefont {Perera},
  \citenamefont {Chamlagain}, \citenamefont {Cheng}, \citenamefont {Yan},
  \citenamefont {Mandrus}, \citenamefont {Tom{\'a}nek},\ and\ \citenamefont
  {Zhou}}]{Chuang:2014}%
  \BibitemOpen
  \bibfield  {author} {\bibinfo {author} {\bibfnamefont {Hsun-Jen}\
  \bibnamefont {Chuang}}, \bibinfo {author} {\bibfnamefont {Xuebin}\
  \bibnamefont {Tan}}, \bibinfo {author} {\bibfnamefont {Nirmal~Jeevi}\
  \bibnamefont {Ghimire}}, \bibinfo {author} {\bibfnamefont
  {Meeghage~Madusanka}\ \bibnamefont {Perera}}, \bibinfo {author}
  {\bibfnamefont {Bhim}\ \bibnamefont {Chamlagain}}, \bibinfo {author}
  {\bibfnamefont {Mark Ming-Cheng}\ \bibnamefont {Cheng}}, \bibinfo {author}
  {\bibfnamefont {Jiaqiang}\ \bibnamefont {Yan}}, \bibinfo {author}
  {\bibfnamefont {David}\ \bibnamefont {Mandrus}}, \bibinfo {author}
  {\bibfnamefont {David}\ \bibnamefont {Tom{\'a}nek}}, \ and\ \bibinfo {author}
  {\bibfnamefont {Zhixian}\ \bibnamefont {Zhou}},\ }\bibfield  {title}
  {\enquote {\bibinfo {title} {High mobility {WSe}$_2$ p- and n-type
  field-effect transistors contacted by highly doped graphene for
  low-resistance contacts},}\ }\href {\doibase 10.1021/nl501275p} {\bibfield
  {journal} {\bibinfo  {journal} {Nano Letters}\ }\textbf {\bibinfo {volume}
  {14}},\ \bibinfo {pages} {3594--3601} (\bibinfo {year} {2014})}\BibitemShut
  {NoStop}%
\bibitem [{\citenamefont {Cui}\ \emph {et~al.}(2015)\citenamefont {Cui},
  \citenamefont {Lee}, \citenamefont {Kim}, \citenamefont {Arefe},
  \citenamefont {Huang}, \citenamefont {Lee}, \citenamefont {Chenet},
  \citenamefont {Zhang}, \citenamefont {Wang}, \citenamefont {Ye},
  \citenamefont {Pizzocchero}, \citenamefont {Jessen}, \citenamefont
  {Watanabe}, \citenamefont {Taniguchi}, \citenamefont {Muller}, \citenamefont
  {Low}, \citenamefont {Kim},\ and\ \citenamefont {Hone}}]{Cui:2015}%
  \BibitemOpen
  \bibfield  {author} {\bibinfo {author} {\bibfnamefont {Xu}~\bibnamefont
  {Cui}}, \bibinfo {author} {\bibfnamefont {Gwan-Hyoung}\ \bibnamefont {Lee}},
  \bibinfo {author} {\bibfnamefont {Young~Duck}\ \bibnamefont {Kim}}, \bibinfo
  {author} {\bibfnamefont {Ghidewon}\ \bibnamefont {Arefe}}, \bibinfo {author}
  {\bibfnamefont {Pinshane~Y.}\ \bibnamefont {Huang}}, \bibinfo {author}
  {\bibfnamefont {Chul-Ho}\ \bibnamefont {Lee}}, \bibinfo {author}
  {\bibfnamefont {Daniel~A.}\ \bibnamefont {Chenet}}, \bibinfo {author}
  {\bibfnamefont {Xian}\ \bibnamefont {Zhang}}, \bibinfo {author}
  {\bibfnamefont {Lei}\ \bibnamefont {Wang}}, \bibinfo {author} {\bibfnamefont
  {Fan}\ \bibnamefont {Ye}}, \bibinfo {author} {\bibfnamefont {Filippo}\
  \bibnamefont {Pizzocchero}}, \bibinfo {author} {\bibfnamefont {Bjarke~S.}\
  \bibnamefont {Jessen}}, \bibinfo {author} {\bibfnamefont {Kenji}\
  \bibnamefont {Watanabe}}, \bibinfo {author} {\bibfnamefont {Takashi}\
  \bibnamefont {Taniguchi}}, \bibinfo {author} {\bibfnamefont {David~A.}\
  \bibnamefont {Muller}}, \bibinfo {author} {\bibfnamefont {Tony}\ \bibnamefont
  {Low}}, \bibinfo {author} {\bibfnamefont {Philip}\ \bibnamefont {Kim}}, \
  and\ \bibinfo {author} {\bibfnamefont {James}\ \bibnamefont {Hone}},\
  }\bibfield  {title} {\enquote {\bibinfo {title} {Multi-terminal transport
  measurements of {MoS}$_2$ using a van der waals heterostructure device
  platform},}\ }\href {\doibase 10.1038/nnano.2015.70} {\bibfield  {journal}
  {\bibinfo  {journal} {Nature Nanotechnology}\ }\textbf {\bibinfo {volume}
  {10}},\ \bibinfo {pages} {534--540} (\bibinfo {year} {2015})}\BibitemShut
  {NoStop}%
\bibitem [{\citenamefont {Liu}\ \emph {et~al.}(2015)\citenamefont {Liu},
  \citenamefont {Wu}, \citenamefont {Cheng}, \citenamefont {Yang},
  \citenamefont {Zhu}, \citenamefont {He}, \citenamefont {Ding}, \citenamefont
  {Li}, \citenamefont {Guo}, \citenamefont {Weiss}, \citenamefont {Huang},\
  and\ \citenamefont {Duan}}]{Liu:2015}%
  \BibitemOpen
  \bibfield  {author} {\bibinfo {author} {\bibfnamefont {Yuan}\ \bibnamefont
  {Liu}}, \bibinfo {author} {\bibfnamefont {Hao}\ \bibnamefont {Wu}}, \bibinfo
  {author} {\bibfnamefont {Hung-Chieh}\ \bibnamefont {Cheng}}, \bibinfo
  {author} {\bibfnamefont {Sen}\ \bibnamefont {Yang}}, \bibinfo {author}
  {\bibfnamefont {Enbo}\ \bibnamefont {Zhu}}, \bibinfo {author} {\bibfnamefont
  {Qiyuan}\ \bibnamefont {He}}, \bibinfo {author} {\bibfnamefont {Mengning}\
  \bibnamefont {Ding}}, \bibinfo {author} {\bibfnamefont {Dehui}\ \bibnamefont
  {Li}}, \bibinfo {author} {\bibfnamefont {Jian}\ \bibnamefont {Guo}}, \bibinfo
  {author} {\bibfnamefont {Nathan~O.}\ \bibnamefont {Weiss}}, \bibinfo {author}
  {\bibfnamefont {Yu}~\bibnamefont {Huang}}, \ and\ \bibinfo {author}
  {\bibfnamefont {Xiangfeng}\ \bibnamefont {Duan}},\ }\bibfield  {title}
  {\enquote {\bibinfo {title} {Toward barrier free contact to molybdenum
  disulfide using graphene electrodes},}\ }\href {\doibase 10.1021/nl504957p}
  {\bibfield  {journal} {\bibinfo  {journal} {Nano Letters}\ }\textbf {\bibinfo
  {volume} {15}},\ \bibinfo {pages} {3030--3034} (\bibinfo {year}
  {2015})}\BibitemShut {NoStop}%
\bibitem [{\citenamefont {Pisoni}\ \emph {et~al.}(2017)\citenamefont {Pisoni},
  \citenamefont {Lee}, \citenamefont {Overweg}, \citenamefont {Eich},
  \citenamefont {Simonet}, \citenamefont {Watanabe}, \citenamefont {Taniguchi},
  \citenamefont {Gorbachev}, \citenamefont {Ihn},\ and\ \citenamefont
  {Ensslin}}]{Prisoni:2017}%
  \BibitemOpen
  \bibfield  {author} {\bibinfo {author} {\bibfnamefont {Riccardo}\
  \bibnamefont {Pisoni}}, \bibinfo {author} {\bibfnamefont {Yongjin}\
  \bibnamefont {Lee}}, \bibinfo {author} {\bibfnamefont {Hiske}\ \bibnamefont
  {Overweg}}, \bibinfo {author} {\bibfnamefont {Marius}\ \bibnamefont {Eich}},
  \bibinfo {author} {\bibfnamefont {Pauline}\ \bibnamefont {Simonet}}, \bibinfo
  {author} {\bibfnamefont {Kenji}\ \bibnamefont {Watanabe}}, \bibinfo {author}
  {\bibfnamefont {Takashi}\ \bibnamefont {Taniguchi}}, \bibinfo {author}
  {\bibfnamefont {Roman}\ \bibnamefont {Gorbachev}}, \bibinfo {author}
  {\bibfnamefont {Thomas}\ \bibnamefont {Ihn}}, \ and\ \bibinfo {author}
  {\bibfnamefont {Klaus}\ \bibnamefont {Ensslin}},\ }\bibfield  {title}
  {\enquote {\bibinfo {title} {Gate-defined one-dimensional channel and broken
  symmetry states in {MoS}$_2$ van der waals heterostructures},}\ }\href
  {\doibase 10.1021/acs.nanolett.7b02186} {\bibfield  {journal} {\bibinfo
  {journal} {Nano Letters}\ }\textbf {\bibinfo {volume} {17}},\ \bibinfo
  {pages} {5008--5011} (\bibinfo {year} {2017})}\BibitemShut {NoStop}%
\bibitem [{\citenamefont {Chee}\ \emph {et~al.}(2019)\citenamefont {Chee},
  \citenamefont {Seo}, \citenamefont {Kim}, \citenamefont {Jang}, \citenamefont
  {Lee}, \citenamefont {Moon}, \citenamefont {Lee}, \citenamefont {Kim},
  \citenamefont {Choi},\ and\ \citenamefont {Ham}}]{Chee:2019}%
  \BibitemOpen
  \bibfield  {author} {\bibinfo {author} {\bibfnamefont {Sang-Soo}\
  \bibnamefont {Chee}}, \bibinfo {author} {\bibfnamefont {Dongpyo}\
  \bibnamefont {Seo}}, \bibinfo {author} {\bibfnamefont {Hanggyu}\ \bibnamefont
  {Kim}}, \bibinfo {author} {\bibfnamefont {Hanbyeol}\ \bibnamefont {Jang}},
  \bibinfo {author} {\bibfnamefont {Seungmin}\ \bibnamefont {Lee}}, \bibinfo
  {author} {\bibfnamefont {Seung~Pil}\ \bibnamefont {Moon}}, \bibinfo {author}
  {\bibfnamefont {Kyu~Hyoung}\ \bibnamefont {Lee}}, \bibinfo {author}
  {\bibfnamefont {Sung~Wng}\ \bibnamefont {Kim}}, \bibinfo {author}
  {\bibfnamefont {Hyunyong}\ \bibnamefont {Choi}}, \ and\ \bibinfo {author}
  {\bibfnamefont {Moon-Ho}\ \bibnamefont {Ham}},\ }\bibfield  {title} {\enquote
  {\bibinfo {title} {Lowering the schottky barrier height by graphene/{Ag}
  electrodes for high-mobility {MoS}$_2$ field-effect transistors},}\ }\href
  {\doibase https://doi.org/10.1002/adma.201804422} {\bibfield  {journal}
  {\bibinfo  {journal} {Advanced Materials}\ }\textbf {\bibinfo {volume}
  {31}},\ \bibinfo {pages} {1804422} (\bibinfo {year} {2019})}\BibitemShut
  {NoStop}%
\bibitem [{\citenamefont {Katoch}\ \emph {et~al.}(2018)\citenamefont {Katoch},
  \citenamefont {Ulstrup}, \citenamefont {Koch}, \citenamefont {Moser},
  \citenamefont {McCreary}, \citenamefont {Singh}, \citenamefont {Xu},
  \citenamefont {Jonker}, \citenamefont {Kawakami}, \citenamefont {Bostwick},
  \citenamefont {Rotenberg},\ and\ \citenamefont {Jozwiak}}]{katoch2018}%
  \BibitemOpen
  \bibfield  {author} {\bibinfo {author} {\bibfnamefont {Jyoti}\ \bibnamefont
  {Katoch}}, \bibinfo {author} {\bibfnamefont {S{\o}ren}\ \bibnamefont
  {Ulstrup}}, \bibinfo {author} {\bibfnamefont {Roland~J.}\ \bibnamefont
  {Koch}}, \bibinfo {author} {\bibfnamefont {Simon}\ \bibnamefont {Moser}},
  \bibinfo {author} {\bibfnamefont {Kathleen~M.}\ \bibnamefont {McCreary}},
  \bibinfo {author} {\bibfnamefont {Simranjeet}\ \bibnamefont {Singh}},
  \bibinfo {author} {\bibfnamefont {Jinsong}\ \bibnamefont {Xu}}, \bibinfo
  {author} {\bibfnamefont {Berend~T.}\ \bibnamefont {Jonker}}, \bibinfo
  {author} {\bibfnamefont {Roland~K.}\ \bibnamefont {Kawakami}}, \bibinfo
  {author} {\bibfnamefont {Aaron}\ \bibnamefont {Bostwick}}, \bibinfo {author}
  {\bibfnamefont {Eli}\ \bibnamefont {Rotenberg}}, \ and\ \bibinfo {author}
  {\bibfnamefont {Chris}\ \bibnamefont {Jozwiak}},\ }\bibfield  {title}
  {\enquote {\bibinfo {title} {Giant spin-splitting and gap renormalization
  driven by trions in single-layer {WS}$_2$/h-{BN} heterostructures},}\ }\href
  {\doibase 10.1038/s41567-017-0033-4} {\bibfield  {journal} {\bibinfo
  {journal} {Nature Physics}\ }\textbf {\bibinfo {volume} {14}},\ \bibinfo
  {pages} {355--359} (\bibinfo {year} {2018})}\BibitemShut {NoStop}%
\bibitem [{\citenamefont {Hinsche}\ \emph {et~al.}(2017)\citenamefont
  {Hinsche}, \citenamefont {Ngankeu}, \citenamefont {Guilloy}, \citenamefont
  {Mahatha}, \citenamefont {Grubi\ifmmode \check{s}\else \v{s}\fi{}i\ifmmode
  \acute{c}\else \'{c}\fi{} \ifmmode~\check{C}\else \v{C}\fi{}abo},
  \citenamefont {Bianchi}, \citenamefont {Dendzik}, \citenamefont {Sanders},
  \citenamefont {Miwa}, \citenamefont {Bana}, \citenamefont {Travaglia},
  \citenamefont {Lacovig}, \citenamefont {Bignardi}, \citenamefont
  {Larciprete}, \citenamefont {Baraldi}, \citenamefont {Lizzit}, \citenamefont
  {Thygesen},\ and\ \citenamefont {Hofmann}}]{Hinsche:2017}%
  \BibitemOpen
  \bibfield  {author} {\bibinfo {author} {\bibfnamefont {Nicki~Frank}\
  \bibnamefont {Hinsche}}, \bibinfo {author} {\bibfnamefont {Arlette~S.}\
  \bibnamefont {Ngankeu}}, \bibinfo {author} {\bibfnamefont {Kevin}\
  \bibnamefont {Guilloy}}, \bibinfo {author} {\bibfnamefont {Sanjoy~K.}\
  \bibnamefont {Mahatha}}, \bibinfo {author} {\bibfnamefont {Antonija}\
  \bibnamefont {Grubi\ifmmode \check{s}\else \v{s}\fi{}i\ifmmode \acute{c}\else
  \'{c}\fi{} \ifmmode~\check{C}\else \v{C}\fi{}abo}}, \bibinfo {author}
  {\bibfnamefont {Marco}\ \bibnamefont {Bianchi}}, \bibinfo {author}
  {\bibfnamefont {Maciej}\ \bibnamefont {Dendzik}}, \bibinfo {author}
  {\bibfnamefont {Charlotte~E.}\ \bibnamefont {Sanders}}, \bibinfo {author}
  {\bibfnamefont {Jill~A.}\ \bibnamefont {Miwa}}, \bibinfo {author}
  {\bibfnamefont {Harsh}\ \bibnamefont {Bana}}, \bibinfo {author}
  {\bibfnamefont {Elisabetta}\ \bibnamefont {Travaglia}}, \bibinfo {author}
  {\bibfnamefont {Paolo}\ \bibnamefont {Lacovig}}, \bibinfo {author}
  {\bibfnamefont {Luca}\ \bibnamefont {Bignardi}}, \bibinfo {author}
  {\bibfnamefont {Rosanna}\ \bibnamefont {Larciprete}}, \bibinfo {author}
  {\bibfnamefont {Alessandro}\ \bibnamefont {Baraldi}}, \bibinfo {author}
  {\bibfnamefont {Silvano}\ \bibnamefont {Lizzit}}, \bibinfo {author}
  {\bibfnamefont {Kristian~Sommer}\ \bibnamefont {Thygesen}}, \ and\ \bibinfo
  {author} {\bibfnamefont {Philip}\ \bibnamefont {Hofmann}},\ }\bibfield
  {title} {\enquote {\bibinfo {title} {Spin-dependent electron-phonon coupling
  in the valence band of single-layer {WS}$_2$},}\ }\href {\doibase
  10.1103/PhysRevB.96.121402} {\bibfield  {journal} {\bibinfo  {journal} {Phys.
  Rev. B}\ }\textbf {\bibinfo {volume} {96}},\ \bibinfo {pages} {121402}
  (\bibinfo {year} {2017})}\BibitemShut {NoStop}%
\bibitem [{\citenamefont {Zhu}\ \emph {et~al.}(2011)\citenamefont {Zhu},
  \citenamefont {Cheng},\ and\ \citenamefont {Schwingenschl\"ogl}}]{Zhu:2011}%
  \BibitemOpen
  \bibfield  {author} {\bibinfo {author} {\bibfnamefont {Z.~Y.}\ \bibnamefont
  {Zhu}}, \bibinfo {author} {\bibfnamefont {Y.~C.}\ \bibnamefont {Cheng}}, \
  and\ \bibinfo {author} {\bibfnamefont {U.}~\bibnamefont
  {Schwingenschl\"ogl}},\ }\bibfield  {title} {\enquote {\bibinfo {title}
  {Giant spin-orbit-induced spin splitting in two-dimensional transition-metal
  dichalcogenide semiconductors},}\ }\href {\doibase
  10.1103/PhysRevB.84.153402} {\bibfield  {journal} {\bibinfo  {journal} {Phys.
  Rev. B}\ }\textbf {\bibinfo {volume} {84}},\ \bibinfo {pages} {153402}
  (\bibinfo {year} {2011})}\BibitemShut {NoStop}%
\bibitem [{\citenamefont {Sio}\ and\ \citenamefont
  {Giustino}(2023)}]{Sio:2023}%
  \BibitemOpen
  \bibfield  {author} {\bibinfo {author} {\bibfnamefont {Weng~Hong}\
  \bibnamefont {Sio}}\ and\ \bibinfo {author} {\bibfnamefont {Feliciano}\
  \bibnamefont {Giustino}},\ }\bibfield  {title} {\enquote {\bibinfo {title}
  {Polarons in two-dimensional atomic crystals},}\ }\href {\doibase
  10.1038/s41567-023-01953-4} {\bibfield  {journal} {\bibinfo  {journal}
  {Nature Physics}\ }\textbf {\bibinfo {volume} {19}},\ \bibinfo {pages}
  {629--636} (\bibinfo {year} {2023})}\BibitemShut {NoStop}%
\bibitem [{\citenamefont {Berkdemir}\ \emph {et~al.}(2013)\citenamefont
  {Berkdemir}, \citenamefont {Guti{\'e}rrez}, \citenamefont
  {Botello-M{\'e}ndez}, \citenamefont {Perea-L{\'o}pez}, \citenamefont
  {El{\'\i}as}, \citenamefont {Chia}, \citenamefont {Wang}, \citenamefont
  {Crespi}, \citenamefont {L{\'o}pez-Ur{\'\i}as}, \citenamefont {Charlier},
  \citenamefont {Terrones},\ and\ \citenamefont {Terrones}}]{Berkdemir:2013}%
  \BibitemOpen
  \bibfield  {author} {\bibinfo {author} {\bibfnamefont {Ayse}\ \bibnamefont
  {Berkdemir}}, \bibinfo {author} {\bibfnamefont {Humberto~R.}\ \bibnamefont
  {Guti{\'e}rrez}}, \bibinfo {author} {\bibfnamefont {Andr{\'e}s~R.}\
  \bibnamefont {Botello-M{\'e}ndez}}, \bibinfo {author} {\bibfnamefont
  {N{\'e}stor}\ \bibnamefont {Perea-L{\'o}pez}}, \bibinfo {author}
  {\bibfnamefont {Ana~Laura}\ \bibnamefont {El{\'\i}as}}, \bibinfo {author}
  {\bibfnamefont {Chen-Ing}\ \bibnamefont {Chia}}, \bibinfo {author}
  {\bibfnamefont {Bei}\ \bibnamefont {Wang}}, \bibinfo {author} {\bibfnamefont
  {Vincent~H.}\ \bibnamefont {Crespi}}, \bibinfo {author} {\bibfnamefont
  {Florentino}\ \bibnamefont {L{\'o}pez-Ur{\'\i}as}}, \bibinfo {author}
  {\bibfnamefont {Jean-Christophe}\ \bibnamefont {Charlier}}, \bibinfo {author}
  {\bibfnamefont {Humberto}\ \bibnamefont {Terrones}}, \ and\ \bibinfo {author}
  {\bibfnamefont {Mauricio}\ \bibnamefont {Terrones}},\ }\bibfield  {title}
  {\enquote {\bibinfo {title} {Identification of individual and few layers of
  {WS}$_2$ using raman spectroscopy},}\ }\href {\doibase 10.1038/srep01755}
  {\bibfield  {journal} {\bibinfo  {journal} {Scientific Reports}\ }\textbf
  {\bibinfo {volume} {3}},\ \bibinfo {pages} {1755} (\bibinfo {year}
  {2013})}\BibitemShut {NoStop}%
\bibitem [{\citenamefont {Novko}(2017)}]{novko_dopant-induced_2017}%
  \BibitemOpen
  \bibfield  {author} {\bibinfo {author} {\bibfnamefont {Dino}\ \bibnamefont
  {Novko}},\ }\bibfield  {title} {\enquote {\bibinfo {title} {Dopant-induced
  plasmon decay in graphene},}\ }\href {\doibase 10.1021/acs.nanolett.7b03553}
  {\bibfield  {journal} {\bibinfo  {journal} {Nano Letters}\ }\textbf {\bibinfo
  {volume} {17}},\ \bibinfo {pages} {6991--6996} (\bibinfo {year}
  {2017})}\BibitemShut {NoStop}%
\bibitem [{\citenamefont {Margine}\ \emph {et~al.}(2016)\citenamefont
  {Margine}, \citenamefont {Lambert},\ and\ \citenamefont
  {Giustino}}]{margine_electron-phonon_2016}%
  \BibitemOpen
  \bibfield  {author} {\bibinfo {author} {\bibfnamefont {E.~R.}\ \bibnamefont
  {Margine}}, \bibinfo {author} {\bibfnamefont {Henry}\ \bibnamefont
  {Lambert}}, \ and\ \bibinfo {author} {\bibfnamefont {Feliciano}\ \bibnamefont
  {Giustino}},\ }\bibfield  {title} {\enquote {\bibinfo {title}
  {Electron-phonon interaction and pairing mechanism in superconducting
  {Ca}-intercalated bilayer graphene},}\ }\href {\doibase 10.1038/srep21414}
  {\bibfield  {journal} {\bibinfo  {journal} {Scientific Reports}\ }\textbf
  {\bibinfo {volume} {6}},\ \bibinfo {pages} {21414} (\bibinfo {year}
  {2016})}\BibitemShut {NoStop}%
\bibitem [{\citenamefont {Bostwick}\ \emph {et~al.}(2010)\citenamefont
  {Bostwick}, \citenamefont {Speck}, \citenamefont {Seyller}, \citenamefont
  {Horn}, \citenamefont {Polini}, \citenamefont {Asgari}, \citenamefont
  {MacDonald},\ and\ \citenamefont {Rotenberg}}]{Bostwick2010}%
  \BibitemOpen
  \bibfield  {author} {\bibinfo {author} {\bibfnamefont {Aaron}\ \bibnamefont
  {Bostwick}}, \bibinfo {author} {\bibfnamefont {Florian}\ \bibnamefont
  {Speck}}, \bibinfo {author} {\bibfnamefont {Thomas}\ \bibnamefont {Seyller}},
  \bibinfo {author} {\bibfnamefont {Karsten}\ \bibnamefont {Horn}}, \bibinfo
  {author} {\bibfnamefont {Marco}\ \bibnamefont {Polini}}, \bibinfo {author}
  {\bibfnamefont {Reza}\ \bibnamefont {Asgari}}, \bibinfo {author}
  {\bibfnamefont {Allan~H.}\ \bibnamefont {MacDonald}}, \ and\ \bibinfo
  {author} {\bibfnamefont {Eli}\ \bibnamefont {Rotenberg}},\ }\bibfield
  {title} {\enquote {\bibinfo {title} {Observation of plasmarons in
  quasi-freestanding doped graphene},}\ }\href {\doibase
  10.1126/science.1186489} {\bibfield  {journal} {\bibinfo  {journal}
  {Science}\ }\textbf {\bibinfo {volume} {328}},\ \bibinfo {pages} {999--1002}
  (\bibinfo {year} {2010})}\BibitemShut {NoStop}%
\bibitem [{\citenamefont {Koppens}\ \emph {et~al.}(2011)\citenamefont
  {Koppens}, \citenamefont {Chang},\ and\ \citenamefont {García~de
  Abajo}}]{koppens_graphene_2011}%
  \BibitemOpen
  \bibfield  {author} {\bibinfo {author} {\bibfnamefont {Frank H.~L.}\
  \bibnamefont {Koppens}}, \bibinfo {author} {\bibfnamefont {Darrick~E.}\
  \bibnamefont {Chang}}, \ and\ \bibinfo {author} {\bibfnamefont {F.~Javier}\
  \bibnamefont {García~de Abajo}},\ }\bibfield  {title} {\enquote {\bibinfo
  {title} {Graphene plasmonics: A platform for strong light–matter
  interactions},}\ }\href {\doibase 10.1021/nl201771h} {\bibfield  {journal}
  {\bibinfo  {journal} {Nano Letters}\ }\textbf {\bibinfo {volume} {11}},\
  \bibinfo {pages} {3370--3377} (\bibinfo {year} {2011})}\BibitemShut {NoStop}%
\bibitem [{\citenamefont {Grigorenko}\ \emph {et~al.}(2012)\citenamefont
  {Grigorenko}, \citenamefont {Polini},\ and\ \citenamefont
  {Novoselov}}]{grigorenko_graphene_2012}%
  \BibitemOpen
  \bibfield  {author} {\bibinfo {author} {\bibfnamefont {A.~N.}\ \bibnamefont
  {Grigorenko}}, \bibinfo {author} {\bibfnamefont {M.}~\bibnamefont {Polini}},
  \ and\ \bibinfo {author} {\bibfnamefont {K.~S.}\ \bibnamefont {Novoselov}},\
  }\bibfield  {title} {\enquote {\bibinfo {title} {Graphene plasmonics},}\
  }\href {\doibase 10.1038/nphoton.2012.262} {\bibfield  {journal} {\bibinfo
  {journal} {Nature Photonics}\ }\textbf {\bibinfo {volume} {6}},\ \bibinfo
  {pages} {749--758} (\bibinfo {year} {2012})}\BibitemShut {NoStop}%
\bibitem [{\citenamefont {Aryasetiawan}\ \emph {et~al.}(1996)\citenamefont
  {Aryasetiawan}, \citenamefont {Hedin},\ and\ \citenamefont
  {Karlsson}}]{aryasetiawan_multiple_1996}%
  \BibitemOpen
  \bibfield  {author} {\bibinfo {author} {\bibfnamefont {F.}~\bibnamefont
  {Aryasetiawan}}, \bibinfo {author} {\bibfnamefont {L.}~\bibnamefont {Hedin}},
  \ and\ \bibinfo {author} {\bibfnamefont {K.}~\bibnamefont {Karlsson}},\
  }\bibfield  {title} {\enquote {\bibinfo {title} {Multiple plasmon satellites
  in {Na} and {Al} spectral functions from ab initio cumulant expansion},}\
  }\href {\doibase 10.1103/PhysRevLett.77.2268} {\bibfield  {journal} {\bibinfo
   {journal} {Physical Review Letters}\ }\textbf {\bibinfo {volume} {77}},\
  \bibinfo {pages} {2268--2271} (\bibinfo {year} {1996})}\BibitemShut {NoStop}%
\bibitem [{\citenamefont {Guzzo}\ \emph {et~al.}(2011)\citenamefont {Guzzo},
  \citenamefont {Lani}, \citenamefont {Sottile}, \citenamefont {Romaniello},
  \citenamefont {Gatti}, \citenamefont {Kas}, \citenamefont {Rehr},
  \citenamefont {Silly}, \citenamefont {Sirotti},\ and\ \citenamefont
  {Reining}}]{guzzo_valence_2011}%
  \BibitemOpen
  \bibfield  {author} {\bibinfo {author} {\bibfnamefont {Matteo}\ \bibnamefont
  {Guzzo}}, \bibinfo {author} {\bibfnamefont {Giovanna}\ \bibnamefont {Lani}},
  \bibinfo {author} {\bibfnamefont {Francesco}\ \bibnamefont {Sottile}},
  \bibinfo {author} {\bibfnamefont {Pina}\ \bibnamefont {Romaniello}}, \bibinfo
  {author} {\bibfnamefont {Matteo}\ \bibnamefont {Gatti}}, \bibinfo {author}
  {\bibfnamefont {Joshua~J.}\ \bibnamefont {Kas}}, \bibinfo {author}
  {\bibfnamefont {John~J.}\ \bibnamefont {Rehr}}, \bibinfo {author}
  {\bibfnamefont {Mathieu~G.}\ \bibnamefont {Silly}}, \bibinfo {author}
  {\bibfnamefont {Fausto}\ \bibnamefont {Sirotti}}, \ and\ \bibinfo {author}
  {\bibfnamefont {Lucia}\ \bibnamefont {Reining}},\ }\bibfield  {title}
  {\enquote {\bibinfo {title} {Valence electron photoemission spectrum of
  semiconductors: Ab initio description of multiple satellites},}\ }\href
  {\doibase 10.1103/PhysRevLett.107.166401} {\bibfield  {journal} {\bibinfo
  {journal} {Physical Review Letters}\ }\textbf {\bibinfo {volume} {107}},\
  \bibinfo {pages} {166401} (\bibinfo {year} {2011})}\BibitemShut {NoStop}%
\bibitem [{\citenamefont {Caruso}\ and\ \citenamefont
  {Giustino}(2016)}]{caruso_gw_2016}%
  \BibitemOpen
  \bibfield  {author} {\bibinfo {author} {\bibfnamefont {Fabio}\ \bibnamefont
  {Caruso}}\ and\ \bibinfo {author} {\bibfnamefont {Feliciano}\ \bibnamefont
  {Giustino}},\ }\bibfield  {title} {\enquote {\bibinfo {title} {The {GW} plus
  cumulant method and plasmonic polarons: application to the homogeneous
  electron gas*},}\ }\href {\doibase 10.1140/epjb/e2016-70028-4} {\bibfield
  {journal} {\bibinfo  {journal} {The European Physical Journal B}\ }\textbf
  {\bibinfo {volume} {89}},\ \bibinfo {pages} {238} (\bibinfo {year}
  {2016})}\BibitemShut {NoStop}%
\bibitem [{\citenamefont {Zhang}\ \emph {et~al.}(2011)\citenamefont {Zhang},
  \citenamefont {Richard}, \citenamefont {Qian}, \citenamefont {Xu},
  \citenamefont {Dai},\ and\ \citenamefont {Ding}}]{CurvM:2011}%
  \BibitemOpen
  \bibfield  {author} {\bibinfo {author} {\bibfnamefont {P.}~\bibnamefont
  {Zhang}}, \bibinfo {author} {\bibfnamefont {P.}~\bibnamefont {Richard}},
  \bibinfo {author} {\bibfnamefont {T.}~\bibnamefont {Qian}}, \bibinfo {author}
  {\bibfnamefont {Y.-M.}\ \bibnamefont {Xu}}, \bibinfo {author} {\bibfnamefont
  {X.}~\bibnamefont {Dai}}, \ and\ \bibinfo {author} {\bibfnamefont
  {H.}~\bibnamefont {Ding}},\ }\bibfield  {title} {\enquote {\bibinfo {title}
  {A precise method for visualizing dispersive features in image plots},}\
  }\href {\doibase 10.1063/1.3585113} {\bibfield  {journal} {\bibinfo
  {journal} {Review of Scientific Instruments}\ }\textbf {\bibinfo {volume}
  {82}},\ \bibinfo {pages} {043712} (\bibinfo {year} {2011})}\BibitemShut
  {NoStop}%
\bibitem [{\citenamefont {Hernangómez-Pérez}\ \emph
  {et~al.}(2023)\citenamefont {Hernangómez-Pérez}, \citenamefont {Donarini},\
  and\ \citenamefont {Refaely-Abramson}}]{hernangomez-perez_charge_2023}%
  \BibitemOpen
  \bibfield  {author} {\bibinfo {author} {\bibfnamefont {Daniel}\ \bibnamefont
  {Hernangómez-Pérez}}, \bibinfo {author} {\bibfnamefont {Andrea}\
  \bibnamefont {Donarini}}, \ and\ \bibinfo {author} {\bibfnamefont {Sivan}\
  \bibnamefont {Refaely-Abramson}},\ }\bibfield  {title} {\enquote {\bibinfo
  {title} {Charge quenching at defect states in transition metal
  dichalcogenide--graphene van der waals heterobilayers},}\ }\href {\doibase
  10.1103/PhysRevB.107.075419} {\bibfield  {journal} {\bibinfo  {journal}
  {Physical Review B}\ }\textbf {\bibinfo {volume} {107}},\ \bibinfo {pages}
  {075419} (\bibinfo {year} {2023})}\BibitemShut {NoStop}%
\bibitem [{\citenamefont {Kresse}\ and\ \citenamefont
  {Hafner}(1993)}]{kresse_ab_1993}%
  \BibitemOpen
  \bibfield  {author} {\bibinfo {author} {\bibfnamefont {G.}~\bibnamefont
  {Kresse}}\ and\ \bibinfo {author} {\bibfnamefont {J.}~\bibnamefont
  {Hafner}},\ }\bibfield  {title} {\enquote {\bibinfo {title} {Ab initio
  molecular dynamics for liquid metals},}\ }\href {\doibase
  10.1103/PhysRevB.47.558} {\bibfield  {journal} {\bibinfo  {journal} {Physical
  Review B}\ }\textbf {\bibinfo {volume} {47}},\ \bibinfo {pages} {558--561}
  (\bibinfo {year} {1993})}\BibitemShut {NoStop}%
\bibitem [{\citenamefont {Kresse}\ and\ \citenamefont
  {Furthmüller}(1996)}]{kresse_efficient_1996}%
  \BibitemOpen
  \bibfield  {author} {\bibinfo {author} {\bibfnamefont {G.}~\bibnamefont
  {Kresse}}\ and\ \bibinfo {author} {\bibfnamefont {J.}~\bibnamefont
  {Furthmüller}},\ }\bibfield  {title} {\enquote {\bibinfo {title} {Efficient
  iterative schemes for ab initio total-energy calculations using a plane-wave
  basis set},}\ }\href {\doibase 10.1103/PhysRevB.54.11169} {\bibfield
  {journal} {\bibinfo  {journal} {Physical Review B}\ }\textbf {\bibinfo
  {volume} {54}},\ \bibinfo {pages} {11169--11186} (\bibinfo {year}
  {1996})}\BibitemShut {NoStop}%
\bibitem [{\citenamefont {Kresse}\ and\ \citenamefont
  {Joubert}(1999)}]{kresse_ultrasoft_1999}%
  \BibitemOpen
  \bibfield  {author} {\bibinfo {author} {\bibfnamefont {G.}~\bibnamefont
  {Kresse}}\ and\ \bibinfo {author} {\bibfnamefont {D.}~\bibnamefont
  {Joubert}},\ }\bibfield  {title} {\enquote {\bibinfo {title} {From ultrasoft
  pseudopotentials to the projector augmented-wave method},}\ }\href {\doibase
  10.1103/PhysRevB.59.1758} {\bibfield  {journal} {\bibinfo  {journal}
  {Physical Review B}\ }\textbf {\bibinfo {volume} {59}},\ \bibinfo {pages}
  {1758--1775} (\bibinfo {year} {1999})}\BibitemShut {NoStop}%
\bibitem [{\citenamefont {Blöchl}(1994)}]{blochl_projector_1994}%
  \BibitemOpen
  \bibfield  {author} {\bibinfo {author} {\bibfnamefont {P.~E.}\ \bibnamefont
  {Blöchl}},\ }\bibfield  {title} {\enquote {\bibinfo {title} {Projector
  augmented-wave method},}\ }\href {\doibase 10.1103/PhysRevB.50.17953}
  {\bibfield  {journal} {\bibinfo  {journal} {Physical Review B}\ }\textbf
  {\bibinfo {volume} {50}},\ \bibinfo {pages} {17953--17979} (\bibinfo {year}
  {1994})}\BibitemShut {NoStop}%
\bibitem [{\citenamefont {Perdew}\ \emph {et~al.}(1996)\citenamefont {Perdew},
  \citenamefont {Burke},\ and\ \citenamefont
  {Ernzerhof}}]{perdew_generalized_1996}%
  \BibitemOpen
  \bibfield  {author} {\bibinfo {author} {\bibfnamefont {John~P.}\ \bibnamefont
  {Perdew}}, \bibinfo {author} {\bibfnamefont {Kieron}\ \bibnamefont {Burke}},
  \ and\ \bibinfo {author} {\bibfnamefont {Matthias}\ \bibnamefont
  {Ernzerhof}},\ }\bibfield  {title} {\enquote {\bibinfo {title} {Generalized
  gradient approximation made simple},}\ }\href {\doibase
  10.1103/PhysRevLett.77.3865} {\bibfield  {journal} {\bibinfo  {journal}
  {Physical Review Letters}\ }\textbf {\bibinfo {volume} {77}},\ \bibinfo
  {pages} {3865--3868} (\bibinfo {year} {1996})}\BibitemShut {NoStop}%
\bibitem [{\citenamefont {Popescu}\ and\ \citenamefont
  {Zunger}(2012)}]{popescu_extracting_2012}%
  \BibitemOpen
  \bibfield  {author} {\bibinfo {author} {\bibfnamefont {Voicu}\ \bibnamefont
  {Popescu}}\ and\ \bibinfo {author} {\bibfnamefont {Alex}\ \bibnamefont
  {Zunger}},\ }\bibfield  {title} {\enquote {\bibinfo {title} {Extracting e vs
  k effective band structure from supercell calculations on alloys and
  impurities},}\ }\href {\doibase 10.1103/PhysRevB.85.085201} {\bibfield
  {journal} {\bibinfo  {journal} {Physical Review B}\ }\textbf {\bibinfo
  {volume} {85}},\ \bibinfo {pages} {085201} (\bibinfo {year}
  {2012})}\BibitemShut {NoStop}%
\bibitem [{\citenamefont {Zheng}(2023)}]{qijingzhengvaspbandunfolding_2023}%
  \BibitemOpen
  \bibfield  {author} {\bibinfo {author} {\bibfnamefont {Qijing}\ \bibnamefont
  {Zheng}},\ }\href {https://github.com/QijingZheng/VaspBandUnfolding}
  {\enquote {\bibinfo {title} {{QijingZheng}/{VaspBandUnfolding}},}\ }
  (\bibinfo {year} {2023}),\ \bibinfo {note} {original-date:
  2017-05-08T08:51:01Z}\BibitemShut {NoStop}%
\bibitem [{\citenamefont {Hofmann}\ \emph {et~al.}(2023)\citenamefont
  {Hofmann}, \citenamefont {Weigl}, \citenamefont {Gradl}, \citenamefont
  {Mishra}, \citenamefont {Orlandini}, \citenamefont {Forti}, \citenamefont
  {Coletti}, \citenamefont {Latini}, \citenamefont {Xian}, \citenamefont
  {Rubio}, \citenamefont {Paredes}, \citenamefont {Causin}, \citenamefont
  {Brem}, \citenamefont {Malic},\ and\ \citenamefont
  {Gierz}}]{hofmann_link_2023}%
  \BibitemOpen
  \bibfield  {author} {\bibinfo {author} {\bibfnamefont {Niklas}\ \bibnamefont
  {Hofmann}}, \bibinfo {author} {\bibfnamefont {Leonard}\ \bibnamefont
  {Weigl}}, \bibinfo {author} {\bibfnamefont {Johannes}\ \bibnamefont {Gradl}},
  \bibinfo {author} {\bibfnamefont {Neeraj}\ \bibnamefont {Mishra}}, \bibinfo
  {author} {\bibfnamefont {Giorgio}\ \bibnamefont {Orlandini}}, \bibinfo
  {author} {\bibfnamefont {Stiven}\ \bibnamefont {Forti}}, \bibinfo {author}
  {\bibfnamefont {Camilla}\ \bibnamefont {Coletti}}, \bibinfo {author}
  {\bibfnamefont {Simone}\ \bibnamefont {Latini}}, \bibinfo {author}
  {\bibfnamefont {Lede}\ \bibnamefont {Xian}}, \bibinfo {author} {\bibfnamefont
  {Angel}\ \bibnamefont {Rubio}}, \bibinfo {author} {\bibfnamefont
  {Dilan~Perez}\ \bibnamefont {Paredes}}, \bibinfo {author} {\bibfnamefont
  {Raul~Perea}\ \bibnamefont {Causin}}, \bibinfo {author} {\bibfnamefont
  {Samuel}\ \bibnamefont {Brem}}, \bibinfo {author} {\bibfnamefont {Ermin}\
  \bibnamefont {Malic}}, \ and\ \bibinfo {author} {\bibfnamefont {Isabella}\
  \bibnamefont {Gierz}},\ }\bibfield  {title} {\enquote {\bibinfo {title} {Link
  between interlayer hybridization and ultrafast charge transfer in
  {WS}$_2$-graphene heterostructures},}\ }\href {\doibase
  10.1088/2053-1583/acdaab} {\bibfield  {journal} {\bibinfo  {journal} {2D
  Materials}\ }\textbf {\bibinfo {volume} {10}},\ \bibinfo {pages} {035025}
  (\bibinfo {year} {2023})}\BibitemShut {NoStop}%
\bibitem [{\citenamefont {Rösner}\ \emph {et~al.}(2015)\citenamefont
  {Rösner}, \citenamefont {Şaşıoğlu}, \citenamefont {Friedrich},
  \citenamefont {Blügel},\ and\ \citenamefont
  {Wehling}}]{rosner_wannier_2015}%
  \BibitemOpen
  \bibfield  {author} {\bibinfo {author} {\bibfnamefont {M.}~\bibnamefont
  {Rösner}}, \bibinfo {author} {\bibfnamefont {E.}~\bibnamefont
  {Şaşıoğlu}}, \bibinfo {author} {\bibfnamefont {C.}~\bibnamefont
  {Friedrich}}, \bibinfo {author} {\bibfnamefont {S.}~\bibnamefont {Blügel}},
  \ and\ \bibinfo {author} {\bibfnamefont {T.~O.}\ \bibnamefont {Wehling}},\
  }\bibfield  {title} {\enquote {\bibinfo {title} {Wannier function approach to
  realistic {Coulomb} interactions in layered materials and
  heterostructures},}\ }\href {\doibase 10.1103/PhysRevB.92.085102} {\bibfield
  {journal} {\bibinfo  {journal} {Physical Review B}\ }\textbf {\bibinfo
  {volume} {92}},\ \bibinfo {pages} {085102} (\bibinfo {year} {2015})},\
  \bibinfo {note} {publisher: American Physical Society}\BibitemShut {NoStop}%
\bibitem [{\citenamefont {Steinke}\ \emph {et~al.}(2020)\citenamefont
  {Steinke}, \citenamefont {Wehling},\ and\ \citenamefont
  {Rösner}}]{steinke_coulomb-engineered_2020}%
  \BibitemOpen
  \bibfield  {author} {\bibinfo {author} {\bibfnamefont {C.}~\bibnamefont
  {Steinke}}, \bibinfo {author} {\bibfnamefont {T.~O.}\ \bibnamefont
  {Wehling}}, \ and\ \bibinfo {author} {\bibfnamefont {M.}~\bibnamefont
  {Rösner}},\ }\bibfield  {title} {\enquote {\bibinfo {title}
  {Coulomb-engineered heterojunctions and dynamical screening in transition
  metal dichalcogenide monolayers},}\ }\href {\doibase
  10.1103/PhysRevB.102.115111} {\bibfield  {journal} {\bibinfo  {journal}
  {Physical Review B}\ }\textbf {\bibinfo {volume} {102}},\ \bibinfo {pages}
  {115111} (\bibinfo {year} {2020})}\BibitemShut {NoStop}%
\bibitem [{\citenamefont {Parcollet}\ \emph {et~al.}(2015)\citenamefont
  {Parcollet}, \citenamefont {Ferrero}, \citenamefont {Ayral}, \citenamefont
  {Hafermann}, \citenamefont {Krivenko}, \citenamefont {Messio},\ and\
  \citenamefont {Seth}}]{parcollet_triqs_2015}%
  \BibitemOpen
  \bibfield  {author} {\bibinfo {author} {\bibfnamefont {Olivier}\ \bibnamefont
  {Parcollet}}, \bibinfo {author} {\bibfnamefont {Michel}\ \bibnamefont
  {Ferrero}}, \bibinfo {author} {\bibfnamefont {Thomas}\ \bibnamefont {Ayral}},
  \bibinfo {author} {\bibfnamefont {Hartmut}\ \bibnamefont {Hafermann}},
  \bibinfo {author} {\bibfnamefont {Igor}\ \bibnamefont {Krivenko}}, \bibinfo
  {author} {\bibfnamefont {Laura}\ \bibnamefont {Messio}}, \ and\ \bibinfo
  {author} {\bibfnamefont {Priyanka}\ \bibnamefont {Seth}},\ }\bibfield
  {title} {\enquote {\bibinfo {title} {{TRIQS}: {A} toolbox for research on
  interacting quantum systems},}\ }\href {\doibase 10.1016/j.cpc.2015.04.023}
  {\bibfield  {journal} {\bibinfo  {journal} {Computer Physics Communications}\
  }\textbf {\bibinfo {volume} {196}},\ \bibinfo {pages} {398--415} (\bibinfo
  {year} {2015})}\BibitemShut {NoStop}%
\bibitem [{\citenamefont {Wentzell}\ \emph {et~al.}(2022)\citenamefont
  {Wentzell}, \citenamefont {Strand}, \citenamefont {Kaeser}, \citenamefont
  {Simon}, \citenamefont {Hampel}, \citenamefont {Egcpvanloon}, \citenamefont
  {Parcollet}, \citenamefont {Philipp},\ and\ \citenamefont
  {Zingl}}]{wentzell_triqstprf_2022}%
  \BibitemOpen
  \bibfield  {author} {\bibinfo {author} {\bibfnamefont {Nils}\ \bibnamefont
  {Wentzell}}, \bibinfo {author} {\bibfnamefont {Hugo~U.R.}\ \bibnamefont
  {Strand}}, \bibinfo {author} {\bibfnamefont {Stefan}\ \bibnamefont {Kaeser}},
  \bibinfo {author} {\bibfnamefont {Dylan}\ \bibnamefont {Simon}}, \bibinfo
  {author} {\bibfnamefont {Alexander}\ \bibnamefont {Hampel}}, \bibinfo
  {author} {\bibnamefont {Egcpvanloon}}, \bibinfo {author} {\bibfnamefont
  {Olivier}\ \bibnamefont {Parcollet}}, \bibinfo {author} {\bibfnamefont
  {D.}~\bibnamefont {Philipp}}, \ and\ \bibinfo {author} {\bibfnamefont
  {Manuel}\ \bibnamefont {Zingl}},\ }\href {\doibase 10.5281/ZENODO.7094480}
  {\enquote {\bibinfo {title} {{TRIQS}/tprf: {Version} 3.1.1},}\ } (\bibinfo
  {year} {2022})\BibitemShut {NoStop}%
\bibitem [{\citenamefont {Giuliani}\ and\ \citenamefont
  {Vignale}(2005)}]{giuliani_quantum_2005}%
  \BibitemOpen
  \bibfield  {author} {\bibinfo {author} {\bibfnamefont {Gabriele}\
  \bibnamefont {Giuliani}}\ and\ \bibinfo {author} {\bibfnamefont {Giovanni}\
  \bibnamefont {Vignale}},\ }\href {\doibase 10.1017/CBO9780511619915} {\emph
  {\bibinfo {title} {Quantum {Theory} of the {Electron} {Liquid}}}}\ (\bibinfo
  {publisher} {Cambridge University Press},\ \bibinfo {address} {Cambridge},\
  \bibinfo {year} {2005})\BibitemShut {NoStop}%
\bibitem [{\citenamefont {Wunsch}\ \emph {et~al.}(2006)\citenamefont {Wunsch},
  \citenamefont {Stauber}, \citenamefont {Sols},\ and\ \citenamefont
  {Guinea}}]{wunsch_dynamical_2006}%
  \BibitemOpen
  \bibfield  {author} {\bibinfo {author} {\bibfnamefont {B}~\bibnamefont
  {Wunsch}}, \bibinfo {author} {\bibfnamefont {T}~\bibnamefont {Stauber}},
  \bibinfo {author} {\bibfnamefont {F}~\bibnamefont {Sols}}, \ and\ \bibinfo
  {author} {\bibfnamefont {F}~\bibnamefont {Guinea}},\ }\bibfield  {title}
  {\enquote {\bibinfo {title} {Dynamical polarization of graphene at finite
  doping},}\ }\href {\doibase 10.1088/1367-2630/8/12/318} {\bibfield  {journal}
  {\bibinfo  {journal} {New Journal of Physics}\ }\textbf {\bibinfo {volume}
  {8}},\ \bibinfo {pages} {318--318} (\bibinfo {year} {2006})}\BibitemShut
  {NoStop}%
\bibitem [{\citenamefont {Hwang}\ and\ \citenamefont
  {Das~Sarma}(2007)}]{hwang_dielectric_2007}%
  \BibitemOpen
  \bibfield  {author} {\bibinfo {author} {\bibfnamefont {E.~H.}\ \bibnamefont
  {Hwang}}\ and\ \bibinfo {author} {\bibfnamefont {S.}~\bibnamefont
  {Das~Sarma}},\ }\bibfield  {title} {\enquote {\bibinfo {title} {Dielectric
  function, screening, and plasmons in two-dimensional graphene},}\ }\href
  {\doibase 10.1103/PhysRevB.75.205418} {\bibfield  {journal} {\bibinfo
  {journal} {Physical Review B}\ }\textbf {\bibinfo {volume} {75}},\ \bibinfo
  {pages} {205418} (\bibinfo {year} {2007})}\BibitemShut {NoStop}%
\bibitem [{\citenamefont {Keldysh}(1979)}]{keldysh_coulomb_1979}%
  \BibitemOpen
  \bibfield  {author} {\bibinfo {author} {\bibfnamefont {L.~V.}\ \bibnamefont
  {Keldysh}},\ }\bibfield  {title} {\enquote {\bibinfo {title} {Coulomb
  interaction in thin semiconductor and semimetal films},}\ }\href@noop {}
  {\bibfield  {journal} {\bibinfo  {journal} {Soviet Journal of Experimental
  and Theoretical Physics Letters}\ }\textbf {\bibinfo {volume} {29}},\
  \bibinfo {pages} {658} (\bibinfo {year} {1979})}\BibitemShut {NoStop}%
\end{thebibliography}
\end{document}